\documentclass[reprint,superscriptaddress,amsmath,amssymb,aps,pra,nofootinbib]{revtex4-1}

\usepackage{graphicx}
\usepackage{dcolumn}
\usepackage{bm}
\usepackage{xcolor}
\usepackage{mathrsfs}
\usepackage{lipsum}
\usepackage{braket}
\usepackage{amsfonts}
\usepackage{hyperref}
\usepackage{cleveref}
\hypersetup{
	pdftitle={},
	colorlinks=true,     
	linkcolor=blue,      
	citecolor=blue,      
	filecolor=black,      
	urlcolor=black        
}

\begin{document}

	\title{Quantum to classical relaxation dynamics of the dissipative Rydberg gas}    
	\author{Viktoria Noel}
    \email[]{viktoria.noel@uni-tuebingen.de}
    \affiliation{Institute for Theoretical Physics, and Center for Integrated Quantum Science and Technology, Universität Tübingen, Auf der Morgenstelle 14, 72076 Tübingen, Germany}
	\author{Igor Lesanovsky}
	\affiliation{Institute for Theoretical Physics, and Center for Integrated Quantum Science and Technology, Universität Tübingen, Auf der Morgenstelle 14, 72076 Tübingen, Germany}
    \affiliation{School of Physics and Astronomy and Centre for the Mathematics and Theoretical Physics of Quantum Non-Equilibrium Systems, University of Nottingham, Nottingham, NG7 2RD, United Kingdom}

	\date{\today}
	
	\begin{abstract}
 
We investigate the relaxation dynamics of a Rydberg gas in regimes where coherent processes and dissipation compete.
In the strongly dissipative limit,  the dynamics is known to be governed by an effective classical rate equation and to exhibit kinetically constrained, glassy relaxation towards a trivial stationary state.
This behaviour originates from the Rydberg blockade, which prevents simultaneous excitations within a characteristic blockade radius.
However, the fate of kinetic constraints in the weakly dissipative limit remains unexplored in large systems above one dimension.
To access large system sizes and two-dimensional geometries, we employ the truncated Wigner approximation, a phase-space method that captures correlated many-body dynamics beyond classical rate equations.
To probe the emergence of kinetic constraints on timescales where coherent and dissipative processes are comparable, we analyse the relaxation dynamics starting from two initial states: a fully polarised state and a Néel state, which belongs to a manifold of so-called quantum scars.
In both cases, we observe a pronounced slowdown in the relaxation of the magnetisation towards the stationary state and identify transient signatures of quantum kinetically constrained dynamics in one and two dimensions. 
		
\end{abstract}

\maketitle

\section{Introduction}
Understanding the relaxation dynamics of open quantum many-body systems remains a central topic of nonequilibrium many-body physics, where the interplay between coherent dynamics and dissipation gives rise to complex behaviour \cite{Weimer_2021, Fazio_2025, Sieberer_2025, Harrington_2022, Diehl_2008, Mi_2024}.
Programmable quantum simulators \cite{Georgescu_2014} based on ultracold atoms \cite{gross2017quantum, bloch2012quantum, Lewenstein_2007} and ions \cite{blatt2012quantum, schneider2012experimental, Monroe_2021}, particularly Rydberg arrays \cite{Weimer_2010, Saffman_2010, L_w_2012, Browaeys_2020}, provide a versatile setting to investigate such dynamics, as long-lived Rydberg states exhibit strong and tunable long-range interactions.
Notably, the strong interactions between Rydberg excitations give rise to a blockade mechanism \cite{urban2009observation, gaetan2009observation, Lukin_2001, Jaksch_2000}, where an excitation suppresses simultaneous excitations within a characteristic blockade radius. 
As a consequence, (de-)excitation events depend on the arrangement of neighbouring excitations, which imposes a kinetic constraint on the many-body dynamics.
This limits the set of accessible configurations and therefore influences relaxation \cite{Dudin_2012, Schau__2012, Barredo_2014}.
Kinetic constraints similar to those found in Rydberg gases have been studied in classical models of glass formers \cite{binder2011glassy, sanders2015sub}.
Here, configuration-dependent transition rules give rise to slow relaxation accompanied by so-called dynamical heterogeneity, i.e., different spatial regions relaxing on substantially different timescales \cite{ritort2003glassy}.
While these models provide valuable insight into the constrained dynamics of classical systems, the investigation of kinetic constraints in the quantum regime remains challenging.
This is especially true for large systems and in higher dimensions, where the exponential growth of the Hilbert space and entanglement generation further complicate the simultaneous treatment of coherent and dissipative processes \cite{Aolita_2015, Fazio_2025, Weimer_2021}.

Approximate descriptions of the many-body dynamics are available in certain regimes.
In the strong dephasing limit, quantum coherences are suppressed, and the evolution can be reduced to an effective classical stochastic process \cite{Lesanovsky_2013, Marcuzzi_2014, Perez_Espigares_2018, Guti_rrez_2015, Everest_2017}.
In this regime, relaxation towards the stationary state proceeds via the aforementioned glassy dynamics characterised by hierarchical relaxation and dynamical heterogeneity.
In the weakly dissipative limit, when quantum coherences become potentially relevant, a description of the dynamics in terms of classical rate equations is generally not applicable. 
In this case, existing studies have been limited to comparatively small system sizes and in one dimension only \cite{Levi_2016, Sch_nleber_2014, Mattioli_2015, Ates_2012, Lesanovsky_2013b, van_Horssen_2015, Lan_2018, rose2020hierarchicalclassicalmetastabilityopen, Olmos_2014, Cech_2025}, with matrix product state simulations enabling the study of larger one-dimensional systems \cite{ cech2025revealingemergentmanybodyphenomena, Klobas_2024, causer2024dynamicalheterogeneitylargedeviations}. 
However, the simulation of dynamics in two spatial dimensions remains difficult due to the substantial computational cost of tensor-network methods in open, long-range interacting systems.
While recent advances in two-dimensional tensor-network and operator-based approaches offer promising routes forward, systematic studies of long-time relaxation remain difficult \cite{Hryniuk_2026, mc2021stable, Kilda_2021, dunham2025efficienttimeevolution2d}.
Therefore, understanding how blockade-induced kinetic constraints manifest in the regime where quantum coherent and dissipative dynamics compete remains an open problem in large, two-dimensional Rydberg gases.

In this work, we investigate the relaxation dynamics in the regime where coherent driving is comparable to or exceeds dissipation.
To access this parameter range for large system sizes and in two dimensions, we employ the truncated Wigner approximation (TWA) \cite{Polkovnikov_2010, polkovnikov2003quantum}.
TWA is a phase-space method that approximates quantum dynamics by an ensemble of classical trajectories, incorporating leading-order quantum fluctuations through statistical sampling of the initial state while remaining computationally tractable beyond exact approaches.
Expectation values of observables, such as magnetisation, as well as correlation functions are obtained as statistical averages over this ensemble of trajectories.
TWA has been widely and successfully applied to isolated quantum many-body systems \cite{blakie2008dynamics, davidson2017semiclassical, Schachenmayer_2015}, and more recently, it has also been increasingly used for dissipative spin systems \cite{Qu_2019, Liu_2020, Huber_2021, Huber_2022, Mink_2022, Singh_2022, Mink_2023, Tebbenjohanns_2024, Zhang:2025kjj, Ruks:2026ndx, Mondal:2026meg, xiang2025quantumpredatorpreycyclesdissipative, Hosseinabadi:2025xbk}.
Such framework is particularly suitable here, as the essential physics of dissipative Rydberg arrays is well captured by interacting spin models with coherent driving and local dissipation.

This paper is organised as follows.
In Sec.~\ref{sec:model} we specify our dissipative spin model under consideration, and in Sec.~\ref{sec:twa} we outline TWA for its dynamics.
In Sec.~\ref{sec:relaxation} we compare TWA to exact dynamics for small system sizes and also to the fully classical description of \cite{Lesanovsky_2013} in the strongly dissipative limit. 
Subsequently, we look at the dynamics for large system sizes in both $d=1,2$, alongside spatial and temporal correlation functions.
We conclude in Sec.~\ref{sec:discussion} with a discussion of the results and an outlook. 

\section{Model}
\label{sec:model}
\begin{figure}
    \centering
\includegraphics[width=1.0\linewidth]{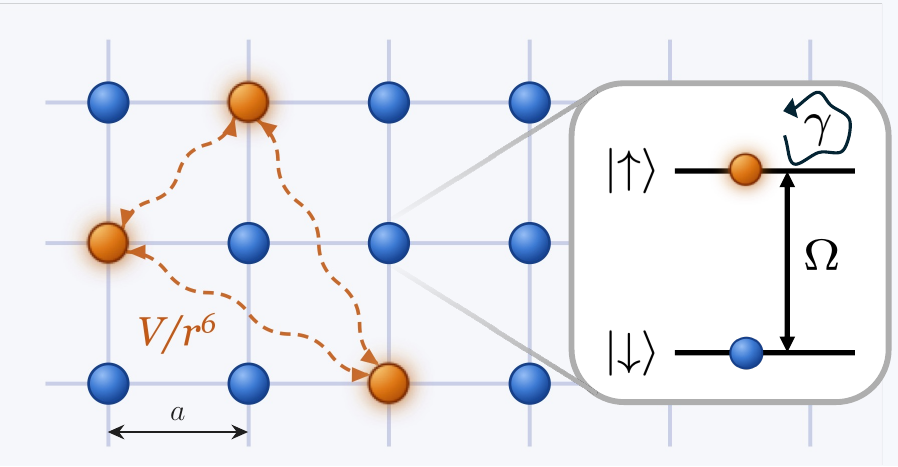}
    \caption{\textbf{Schematic representation of interacting two-level atoms on a two-dimensional lattice.} The ground state $\ket{\downarrow}$ (blue) is coupled to the Rydberg state $\ket{\uparrow}$ (orange) by a laser with Rabi frequency $\Omega$. Atoms in the Rydberg state interact via the potential $V/r^6$, with $r$ being the interatomic separation. Throughout, we measure distances, such as the interparticle separation $|\mathbf{r}_i-\mathbf{r}_j|$, in units of lattice spacing $a$.
    }
    \label{fig:schematic}
\end{figure}

The essential physics of Rydberg systems is captured by paradigmatic quantum spin models, for instance the long-range interacting transverse-field Ising model.
We consider a system of $N$ atoms at the sites of a regular lattice in $d=1$ and $d=2$ dimensions with uniform lattice spacing $a$ and periodic boundary conditions. 
Each atom is modelled as a two-level system with a ground state $\ket{\downarrow}$ and a highly excited Rydberg state $\ket{\uparrow}$, as schematically shown in Fig.~\ref{fig:schematic} for a two-dimensional lattice.
At each lattice site $i$, we can map this onto a quantum spin-$1/2$ particle. 
In this representation, the states $\ket{\downarrow}$ and $\ket{\uparrow}$ correspond to the eigenstates of the Pauli operator $\hat{\sigma}_i^z$, and we introduce the usual spin operators $\hat{\sigma}_i^{x,y,z}$ acting on site $i$. 
The local Rydberg excitation number operator is given by
\begin{equation}
\hat{n}_i = \frac{1 + \hat{\sigma}_i^z}{2},
\label{eq:exc-density}
\end{equation}
which projects onto the excited state.
We consider two different initial states. 
The first is the fully polarised state
\begin{equation}
|\Psi_{\downarrow}(0)\rangle
=
\bigotimes_{i=1}^N \ket{\downarrow}_i ,
\label{eq:ic}
\end{equation}
which is the natural initial condition for the relaxation dynamics since all spins start in the ground state.
The second one is the Néel state
\begin{equation}
|\Psi_{\mathrm{N\acute{e}el}}(0)\rangle
=
\bigotimes_{i\in A}\ket{\uparrow}_i
\bigotimes_{j\in B}\ket{\downarrow}_j ,
\label{eq:ic2}
\end{equation}
which represents a staggered configuration on the two sublattices.
In the limit of strong interactions and in the absence of dissipation, this state exhibits quantum scarring: it belongs to a low-dimensional manifold of states and the ensuing dynamics displays persistent coherent oscillations \cite{Bernien_2017, Turner_2018, Serbyn_2021}. 
Considering this initial state therefore allows to illuminate the role of quantum coherence during relaxation.

When coupled to an external environment, the evolution of the system's density matrix is described by the master equation
\begin{equation}
\frac{d}{d t} \hat{\rho}\equiv\mathcal{L}[\hat{\rho}]=-i[\hat{H}, \hat{\rho}]+\mathcal{D}[\hat{\rho}].
\label{eq:Lindblad}
\end{equation}
Here, the first term describes the coherent evolution due to the Hamiltonian $\hat{H}$, and the subsequent dissipator $\mathcal{D}$ describes the incoherent processes. 
Each dissipator is associated with a specific jump process, 
\begin{equation}
\mathcal{D}[\hat{\rho}]=\sum_i\left(\hat{L}_i \hat{\rho} \hat{L}_i^{\dagger}-\frac{1}{2}\left\{\hat{L}_i^{\dagger} \hat{L}_i, \hat{\rho}\right\}\right),
\end{equation}
described by a set of jump operators $\hat{L}_i$. 
The coherent part of the system's evolution is governed by a Hamiltonian, 
\begin{equation}
\hat{H} = \Omega \sum_{i=1}^{N} \hat{\sigma}_i^x + \sum_{i<j} V_{ij} \hat{n}_i \hat{n}_j,
\end{equation}
where an external laser field couples the ground and Rydberg states, and the Rydberg-Rydberg interactions are described by a long-range potential, with 
\begin{equation}
V_{ij}=\frac{V} {\left|\mathbf{r}_i-\mathbf{r}_j\right|^\alpha}.
\label{eq:longrange}
\end{equation}
In the following we consider van der Waals interactions with $\alpha=6$, and we regard dephasing as the primary source of dissipation.
In practice this arises from environmental noise sources, such as laser phase fluctuations or thermal motion, which randomise the relative phase between the ground and Rydberg states, effectively destroying quantum coherence \cite{Saffman_2010}. 
At site $i$, the jump operator is 
\begin{equation}
\hat{L}_{i, \gamma}=\sqrt{\gamma} \hat{n}_i,
\end{equation}
where $\gamma$ is the dephasing rate. 

\section{Truncated Wigner approach}
\label{sec:twa}
To access system sizes beyond the reach of exact methods, particularly in two spatial dimensions, we employ the truncated Wigner approximation.
In this approach, the quantum master equation is mapped onto a stochastic evolution of classical phase-space variables, which can be obtained from the Keldysh path-integral representation \cite{Hosseinabadi:2025xbk}. 
A detailed derivation of this mapping and that of the TWA equations used in this work is provided in App.~\ref{app:TWA}, here we summarise the general form of the resulting evolution equations.
In essence, operator-valued degrees of freedom $\hat{\psi}_{\alpha}$, corresponding in our case to the Pauli operators $\sigma_i^{\alpha}$ with $\alpha = x,y,z$, are replaced by classical phase-space variables $\psi_{\alpha} \equiv s_i^{\alpha}$.
Similarly, the Hamiltonian and jump operators are mapped to their classical counterparts $\hat{H} \rightarrow H, \hat{L} \rightarrow L$, yielding equations of motion that approximate the quantum dynamics at leading order in quantum fluctuations \cite{Polkovnikov_2010}.
For an open system with Hamiltonian $H$ and jump operators $L_i$, the resulting stochastic evolutions take the form
\begin{equation}
\frac{d}{d t} \psi_\alpha=\left\{\psi_\alpha, H\right\}_p-\frac{i}{2} \sum_i\left(\left\{\psi_\alpha, L^*_i\right\}_p \Phi_i-\Phi^*_i\left\{L_i, \psi_\alpha\right\}_p\right),
\label{eq:twa-stoch}
\end{equation}
with the Poisson bracket $\{\cdot,\cdot\}_p$. 
In practice, observables are obtained by evolving an ensemble of classical trajectories according to Eq.~\eqref{eq:twa-stoch}, with initial conditions sampled from a phase-space representation of the quantum state. 
The corresponding classical observable is evaluated along each stochastic trajectory and averaged over the ensemble. 
This procedure approximates the expectation values computed from the density matrix evolved with the Lindblad equation \eqref{eq:Lindblad}.
The fields $\Phi_i$ appearing in \eqref{eq:twa-stoch} are defined as 
\begin{equation}
\Phi_i = \gamma L_i + \xi_i(t),
\label{eq:auxiliaryfields}
\end{equation} 
where $\xi_i(t)$ are complex Gaussian noise fields with correlations \begin{equation}
\overline{\xi_i(t)\xi^*_j(t')} = 2\gamma \delta_{ij}\delta(t-t'),
\end{equation}
in this case for local dephasing at rate $\gamma$. 
The overline in the above equation denotes the average over noise realisations.
The Kronecker delta $\delta_{ij}$ implies that the noise is local in space such that fluctuations associated with different jump operators are uncorrelated. 
The Dirac delta $\delta(t-t')$ corresponds to Markovian white noise in time. 
More general noise structures, including correlated noise, can be incorporated by replacing these delta functions with an appropriate correlation kernel.

In the discrete truncated Wigner approximation (DTWA) \cite{wootters1987wigner,
Schachenmayer_2015, Sundar_2019, zhu2019generalized}, the initial phase-space distribution is replaced by a discrete sampling scheme.
For the fully polarised state \eqref{eq:ic}, all spins are aligned along the negative $z$-direction, such that 
$\langle \hat{\sigma}_k^z \rangle = -1$ and $\langle \hat{\sigma}_k^{x,y} \rangle = 0$.
For the Néel initial state \eqref{eq:ic2}, spins on the two sublattices are polarised in opposite directions, with 
$\langle \hat{\sigma}_i^z \rangle = +1$ for $i\in A$ and $\langle \hat{\sigma}_j^z \rangle = -1$ for $j\in B$.
In both cases the transverse spin components have vanishing expectation values, while their fluctuations are maximal,
$\langle (\hat{\sigma}_k^x)^2 \rangle = \langle (\hat{\sigma}_k^y)^2 \rangle = 1$.
Accordingly, the classical spin variables $s_i^\alpha$ are sampled from the discrete distribution
\begin{equation}
W_0(s_i^x,s_i^y,s_i^z)
=
\begin{cases}
\frac{1}{4}, & (s_i^x,s_i^y,s_i^z)=(\pm1,\pm1,s_i^z), \\
0, & \text{otherwise},
\end{cases}
\end{equation}
where $s_i^z=-1$ for the fully polarised state and $s_i^z=+1$ ($-1$) for sites on sublattice $A$ ($B$) in the Néel state.
This sampling exactly reproduces the first and second moments of the corresponding quantum spin operators.
The sampled spin configurations are subsequently evolved according to the stochastic classical equations of motion (see App.~\ref{app:TWA}), given by
\begin{align}
\frac{d}{dt} s_i^x
&= - V_i^{\mathrm{eff}} s_i^y
+ 2\eta_i(t)\, s_i^y,
\\[6pt]
\frac{d}{dt} s_i^y
&= V_i^{\mathrm{eff}} s_i^x
- 2\Omega s_i^z
- 2\eta_i(t)\, s_i^x,
\\[6pt]
\frac{d}{dt} s_i^z
&= 2\Omega\, s_i^y,
\end{align}
where
\begin{align}
V_i^{\mathrm{eff}}
&= \sum_{j\neq i}
\frac{V_{ij}}{2}\left(1+s_j^z\right),
\label{eq:veff}
\end{align}
and $V_{ij}$ is given by Eq.~\eqref{eq:longrange}. 
For local dephasing with the Hermitian jump operator $L_i=\sqrt{\gamma}\,n_i$, the complex noise fields $\xi_i(t)$ can be recast as a single real noise field $\eta_i(t)$ acting on the transverse spin components, as only the combination $\eta_i(t) \equiv (\xi_i(t)+\bar{\xi}_i(t))/2$ appears in Eq.~\eqref{eq:twa-stoch}, giving
\begin{equation}
\overline{\eta_i(t)\eta_{j}(t')} = \gamma\,\delta_{ij}\delta(t-t').
\label{eq:realnoise}
\end{equation}

The stochastic equations are interpreted in the Stratonovich sense, consistent with their derivation from the Keldysh path-integral formalism \cite{Hosseinabadi:2025xbk}. 
For the numerical integration, we employ a rotation-based integrator that evolves each spin by a finite element of $\mathrm{SO}(3)$. 
In this scheme, the deterministic and stochastic contributions define an effective field about which the spin precesses, and the discrete update is implemented as the corresponding finite rotation on the Bloch sphere.
This construction preserves the spin length $|\mathbf s_i|$ exactly and ensures consistency with Stratonovich calculus. 
Throughout this work we express all lengths in units of the lattice spacing $a$ and time is expressed in units of $\Omega^{-1}$.
The time stepping used in the simulations is $\Delta t=10^{-5}$.
The numerical implementation is described in detail in App.~\ref{app:numerics}.
\begin{figure*}[htb]
    \centering
    \includegraphics[width=0.99\linewidth]{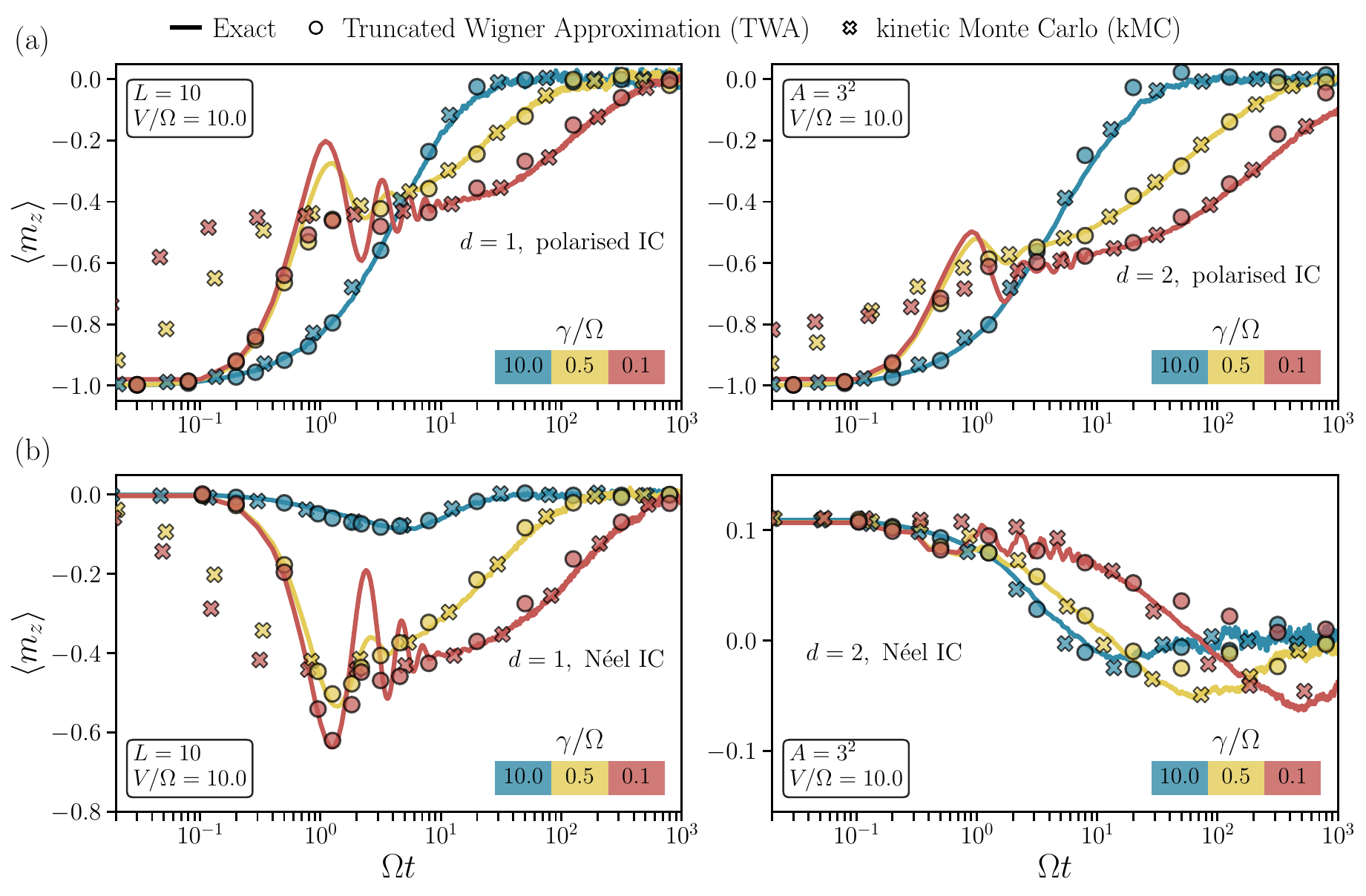}
    \caption{\textbf{Benchmarking the TWA against exact and kMC dynamics for different dissipative strengths.} The magnetisation relaxation from the fully polarised initial state (a) as well as the Néel state (b) is shown for different dephasing $\gamma / \Omega=10.0, 0.5,0.1$ and interaction $V/ \Omega=10.0$ in $d=1$ (left) and $d=2$ (right). The corresponding strongly dissipative (rate-equation) blockade radii are $R=0.89, 1.47, 1.92$ for decreasing $\gamma / \Omega.$ For the large dephasing limit $\gamma / \Omega = 10.0$, the exact, kMC and TWA data all agree. For lower dephasing $\gamma/ \Omega=0.1,0.5$, the early to intermediate dynamics qualitatively differ, including the presence of some coherent oscillations, whose amplitude TWA underestimates at this interaction strength. However, the long-time behaviour agrees for all three methods, with slight deviations in $d=2$ for $\gamma / \Omega = 0.1$.}
    \label{fig:bench1}
\end{figure*}

\section{Relaxation dynamics}
\label{sec:relaxation}
\subsection{Benchmark and kinetic constraints in small systems}
\label{sec:bench}
\begin{figure*}[htb]
    \centering
    \includegraphics[width=0.99\linewidth]{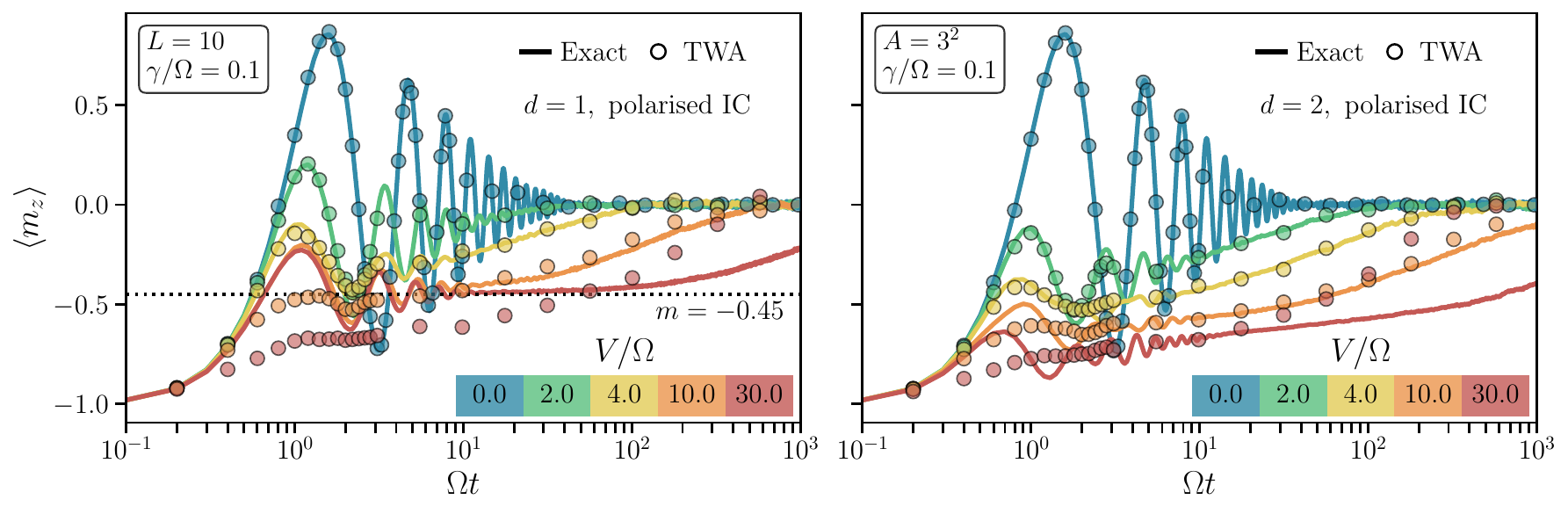}
    \caption{\textbf{Benchmarking the TWA against exact dynamics for different interaction strengths at low dephasing.} The magnetisation relaxation is shown for different interactions $V/\Omega = 0.0-30.0$ and $\gamma/\Omega=0.1$ for $d=1$ (left) and $d=2$ (right). TWA accurately reproduces the relaxation dynamics from early to late times, and up to $V/\Omega \sim 5.0$, also qualitatively captures the intermediate-time oscillatory behaviour. Above $V/\Omega = 10.0$, TWA can no longer qualitatively describe the time evolution of the magnetisation, in particular it fails to reproduce the intermediate plateau at $m\sim-0.45$ in $d=1$.}
    \label{fig:bench2}
\end{figure*}
We first benchmark the TWA approach against numerically exact simulations obtained via stochastic unravelling of the Lindblad master equation based on the quantum-jump Monte Carlo method \cite{Daley_2014}.
This allows us to test how well TWA captures quantum coherent effects in the relaxation dynamics.
To this end, we consider the relaxation of the mean magnetisation,
\begin{equation}
\langle m_z(t) \rangle
= \frac{1}{N} \sum_{i=1}^{N} \langle s_i^z(t) \rangle,
\end{equation}
where the expectation value is obtained by averaging over an ensemble of quantum trajectories in the stochastic unravelling.
The quantum-jump simulations are performed using the package QuantumToolbox.jl \cite{Mercurio_2025}.
In TWA, the same observable is evaluated by averaging over ensembles of classical trajectories.
In this work, we average over 5000 trajectories for one-dimensional, and 3000 trajectories for two-dimensional results.
Beyond mean values, the ensemble directly provides the uncertainty of each
observable: the standard deviation across trajectories, $\sigma_O = \sqrt{\langle O^2\rangle - \langle O\rangle^2}$, which estimates the
physical shot-to-shot fluctuations, while the statistical error of the mean is $\sigma_O/\sqrt{M}$. 
For the mean magnetisation this error is of order $10^{-3}$ for the trajectory numbers used here, i.e.\ smaller than the linewidths in the figures shown below.

\subsubsection{Coherent and classical relaxation regimes}
A particularly interesting regime arises when the dephasing rate is much larger than the Rabi frequency, $\gamma \gg \Omega$. 
In this strongly dissipative limit, coherences decay rapidly and can be adiabatically eliminated. 
The dynamics is then described by an effective classical master equation \cite{Lesanovsky_2013}
\begin{equation}
\partial_t \rho=\sum_i \Gamma_i\left(\sigma_i^{+}\rho\sigma_i^{-}+\sigma_i^{-}\rho\sigma_i^{+}-\rho\right).
\label{eq:rate-eq}
\end{equation}
This describes stochastic spin flips $\ket{\downarrow}_i \leftrightarrow \ket{\uparrow}_i$ with configuration-dependent rates $\Gamma_i$ given by
\begin{equation}
\Gamma_i=\frac{4 \Omega^2}{\gamma} \frac{1}{1+\left(R^\alpha \sum_{j \neq i} \frac{n_j}{\left|\mathbf{r}_j-\mathbf{r}_i\right|^\alpha}\right)^2},
\label{eq:rates}
\end{equation}
with the interaction parameter $R=\left( V / 2\gamma\right)^{1 / \alpha}$, where $V$ is given by Eq.~\eqref{eq:longrange}.
The resulting stochastic dynamics can be simulated efficiently using kinetic Monte Carlo (kMC) as studied in \cite{Lesanovsky_2013}, allowing access to much larger system sizes than in the full quantum evolution.
The validity of this approach has also been confirmed experimentally \cite{Schempp_2014, Urvoy_2015, Valado_2016}.

Representative data for a spin chain of length $L=10$, and a spin lattice of $A=3^2$ with interaction strength $V/\Omega=10$ are shown for the fully polarised as well as the Néel initial states in Fig.~\ref{fig:bench1}(a) and (b), respectively.
In the limit of strong dephasing ($\gamma/ \Omega=10.0$, blue curve), the exact, TWA and kMC relaxation curves agree for all timescales for both initial conditions.
For smaller dephasing rates ($\gamma /\Omega=0.5, 0.1$, yellow and red curves), the dynamics changes qualitatively and coherent oscillations are present at early to intermediate time scales.
For the Néel initial state these oscillations originate from the quantum many-body scar manifold, which identifies the time window where coherent quantum dynamics dominates before noise-driven relaxation takes over.
In this regime, kMC systematically deviates from the exact quantum dynamics, as Eq.~\eqref{eq:rate-eq} is no longer applicable.
On the other hand, TWA remains in good qualitative and quantitative agreement with the exact results at early and late times, although it underestimates the amplitude of the intermediate oscillations.
We emphasise that the coherent oscillations are not a perturbative correction to the classical dynamics, but rather, they reflect a regime where coherent many-body dynamics governs the evolution, inaccessible to classical rate equations by construction.

However, at long times, the final stage of relaxation is governed by an exponential decay towards the stationary state, which is well captured by both TWA and kMC for all values of $\gamma$, even though the classical rate equation underlying kMC is formally derived only for $\gamma \gg \Omega$.
In this late-time regime, the system predominantly occupies configurations with magnetisation close to zero, which are strongly interacting and individual spin flips involve a large interaction-induced energy cost or gain. 
These transitions are therefore strongly off-resonant and cannot be efficiently driven by coherent processes, but instead rely on dissipative spin flips. 
Consequently, the dynamics is governed by slow, dissipative configuration changes, which are well captured by the classical rate equation \eqref{eq:rate-eq}, even when coherent processes affect the intermediate-time behaviour.

The relaxation dynamics therefore involves a crossover from an early regime dominated by coherent many-body dynamics to a late-time regime where dissipation governs the slow approach to the stationary state.
As seen in the next subsection, strong interactions also impose kinetic constraints that restrict the allowed configuration changes.
Understanding this crossover and the role of kinetic constraints in the weakly dissipative scenario is a central aspect of the behaviour studied here.

\subsubsection{Emergence of kinetic constraints at weak dissipation}
Fig.~\ref{fig:bench2} shows the relaxation of $\langle m_z(t) \rangle$ for increasing interaction strength $V/\Omega$ in $d=1,2$ dimensions for the fully polarised initial state.
For weak interactions, $V/\Omega \lesssim 5.0$, TWA quantitatively reproduces the exact dynamics for all times, while for stronger interactions, $V/\Omega \lesssim 10.0$, the amplitude of intermediate oscillations is gradually underestimated.
For larger interaction strengths, $V/\Omega \gtrsim 10.0$, there is a qualitative disagreement between TWA and the exact dynamics.
Here, the relaxation is increasingly influenced by kinetic constraints, which originate from the interplay between coherent laser excitation and strong interactions between excited atoms.
In the absence of interactions, the drive $\Omega$ induces resonant spin flips between the ground and excited states.
However, when an atom is already excited, it shifts the energy levels of nearby atoms via the van der Waals interaction,  rendering the excitation of a neighbouring atom off-resonant when $V_{ij} \gg \Omega$ \cite{Lesanovsky_2013}, a manifestation of the Rydberg blockade.
As a result, the local configuration determines whether a transition is energetically accessible, imposing an effective kinetic constraint on the dynamics. 
Since the van der Waals interaction decays as $V_{ij} \sim 1/r_{ij}^6$, nearest-neighbour interactions strongly exceed the coherent drive, $V_{\mathrm{NN}} \gg \Omega$, while next-nearest-neighbour contributions can become comparable to $\Omega$. 
Consequently, spin flips in the immediate vicinity of an excitation are strongly off-resonant.
This kinetic constraint manifests in two ways: the emergence of a transient plateau in the magnetisation at intermediate times, and a slow, dissipation-dominated final approach to the stationary state.

For the fully polarised initial state in one dimension, this occurs at $\langle m_z\rangle \simeq -0.45$ in the exact numerics.
This value corresponds to that of a one-dimensional lattice gas subject to a hard-dimer constraint, where the presence of an excitation on a site suppresses the excitation of its nearest neighbours \cite{Lesanovsky_2011, Ji_2013, Olmos_2010}.
The magnetisation plateau thus suggests that the Rydberg blockade effectively imposes a nearest-neighbour exclusion constraint on the dynamics: once the system reaches a configuration where the simultaneous excitation of neighbouring sites are blocked, further relaxation is strongly suppressed.
While TWA does not capture the extended plateau arising from this at $V/\Omega=30.0$, it does reproduce the key signatures of kinetic constraints at $V/\Omega=10.0$: the onset of a plateau in the magnetisation and the slowed, dissipation-dominated final relaxation, as seen in Fig.~\ref{fig:bench1}(a) and Fig.~\ref{fig:bench2}.
For the Néel initial state, no intermediate plateau develops as seen in Fig.~\ref{fig:bench1}(b).
In contrast to the fully polarised case, this configuration already contains  excitations on alternating sites, so the relaxation proceeds mainly through their rearrangement rather than through a gradual increase in the excitation density that leads to the intermediate plateau.
For weaker dissipation, resonant transitions lead to pronounced oscillations, causing the magnetisation to develop a short-lived minimum at intermediate times before slowly relaxing towards the stationary state.
At even stronger interactions, the dynamics is effectively captured by a PXP model with dephasing in which Hilbert-space fragmentation gives rise to a large set of long-lived metastable states fixed by the initial configuration \cite{yan2025}. 
For strong dissipation, the magnetisation changes only slightly during the evolution (Fig.~\ref{fig:bench1}(b), blue curve). 
This behaviour is consistent with relaxation dominated by noise-induced off-resonant processes.

These observations confirm a clear separation between the quantum and classical regimes at weak dissipation and the role of kinetic constraints.
At early times, the dynamics is dominated by coherent, resonant transitions for both initial conditions.
For the fully polarised initial state, this regime leads to the emergence of the blockade-induced plateau discussed above. 
The classical rate equation overestimates its extent (red curve in Fig.~\ref{fig:bench1}(a) for $d=1$) because it cannot capture the coherent oscillations present at weak dissipation. 
By contrast, for the Néel initial state the system does not pass through such an intermediate jammed configuration.
Instead, coherent effects, that are not captured by the classical rate equation lead to pronounced oscillations that can reach below the minimum predicted by the classical rate equation, before relaxing to the same stationary state.

For the two-dimensional $A=3^2$ lattice, the relaxation dynamics differs qualitatively from the one-dimensional $L=10$ case. 
For the polarised initial state, only the onset of a plateau is visible, which becomes more pronounced with both increasing interaction strength and as shown in the following section, with system size.
This indicates that, while local blockade effects still suppress nearby excitations, they do not generate a sufficiently sharp separation of time scales to stabilise a long-lived plateau.
However, the onset of the plateau observed here is itself a signature of quantum-coherent dynamics, as no such feature was observed in the strongly dissipative regime \cite{Lesanovsky_2013} in $d=2$.
In the weakly dissipative case, coherent dynamics gives rise to this plateau, followed by a slower, dissipation-dominated relaxation.
The Néel initial state in $d=2$ suffers from finite-size effects for the small $3^2$ lattice.
To investigate this further, we apply TWA in the following to larger systems.

\subsection{Kinetic constraints in large systems}
A key advantage of TWA is its favourable scaling with system size compared to exact quantum simulations, where the Hilbert space grows exponentially.
We therefore make use of this scalability to investigate substantially larger systems beyond the reach of numerically exact stochastic unravellings of the Lindblad equation.
In one dimension, we consider spin chains of length $L=100$, while in two dimensions we study square lattices of size $A=15^2$. 
We focus on the regime $V/\Omega \leq 10.0$ and $\gamma/\Omega \geq 0.1$, where TWA was found to reproduce the exact quantum dynamics for small system sizes.

\subsubsection{Magnetisation relaxation}
\begin{figure*}
    \centering
    \includegraphics[width=0.99\linewidth]{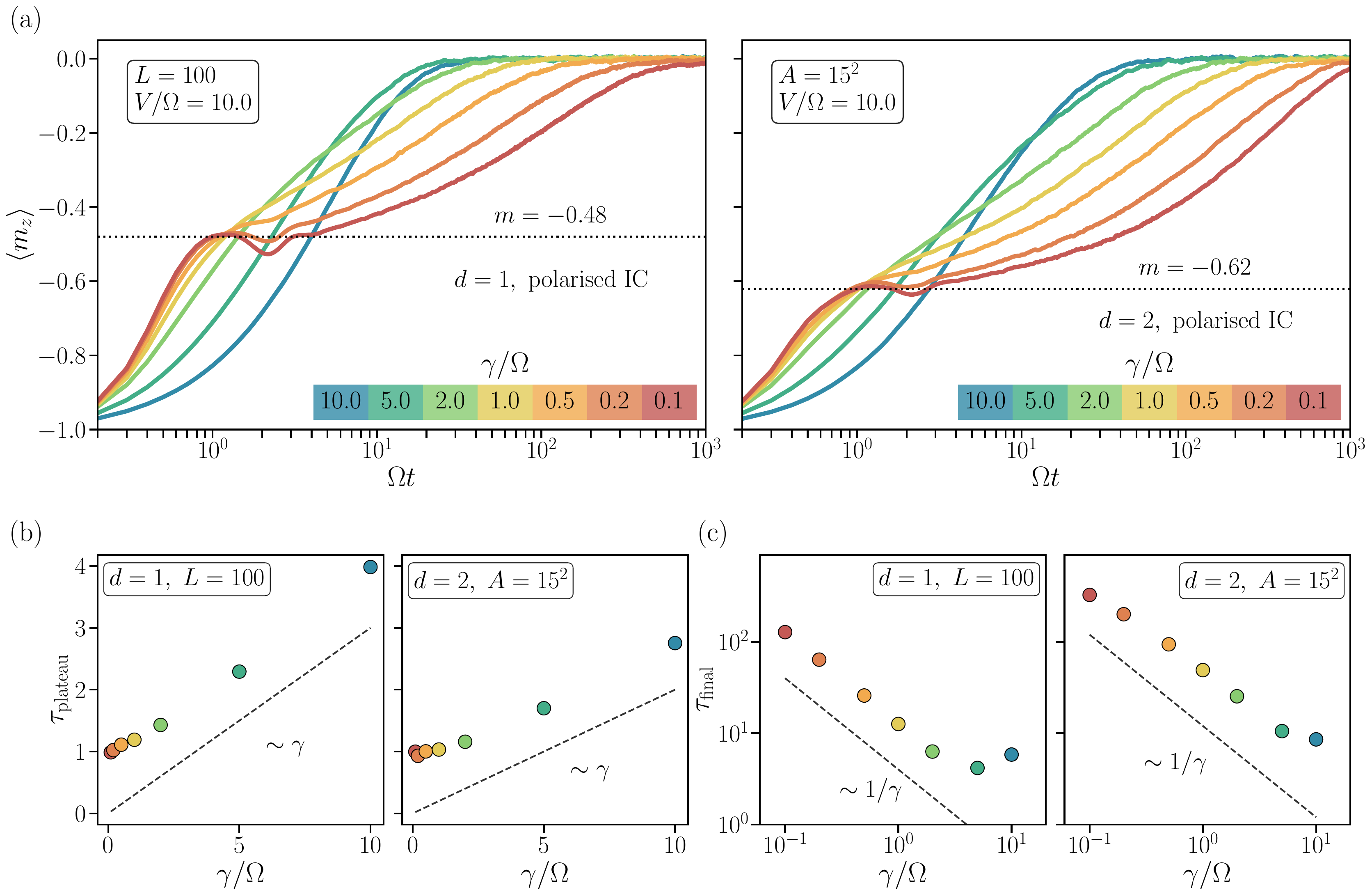}
    \caption{\textbf{Magnetisation relaxation at strong interaction and different dephasing strengths for the fully polarised initial state.}
    (a) Time evolution of the magnetisation for $V/\Omega=10.0$ in $d=1$ (left) and $d=2$ (right). 
    As $\gamma/\Omega$ decreases, the relaxation develops an intermediate plateau at $\langle m_z\rangle \simeq -0.48$ in $d=1$ and $\langle m_z\rangle \simeq -0.62$ in $d=2$, reflecting the emergence of interaction-induced kinetic constraints.
    (b) Plateau time $\tau_{\text{plateau}}$ (left), defined as the time at which $\langle m_z(t)\rangle=-0.48$ is reached in $d=1$ and $\langle m_z(t)\rangle=-0.62$ in $d=2$. 
    The approach to the plateau scales approximately linearly with $\gamma$, indicating that stronger dephasing suppresses coherent processes and slows the buildup of excitations.
    (c) The final relaxation time extracted from an exponential fit shows a scaling $\tau \sim 1/\gamma$ for $\gamma \lesssim \Omega$.
    The deviation at the largest dephasing strengths reflects the crossover to the quantum Zeno regime, where strong dephasing suppresses coherent flips and $\tau$ instead grows as $\sim \gamma/\Omega^2$.}
    \label{fig:mag_rel_taus}
\end{figure*}

\begin{figure*}
    \centering
    \includegraphics[width=0.99\linewidth]{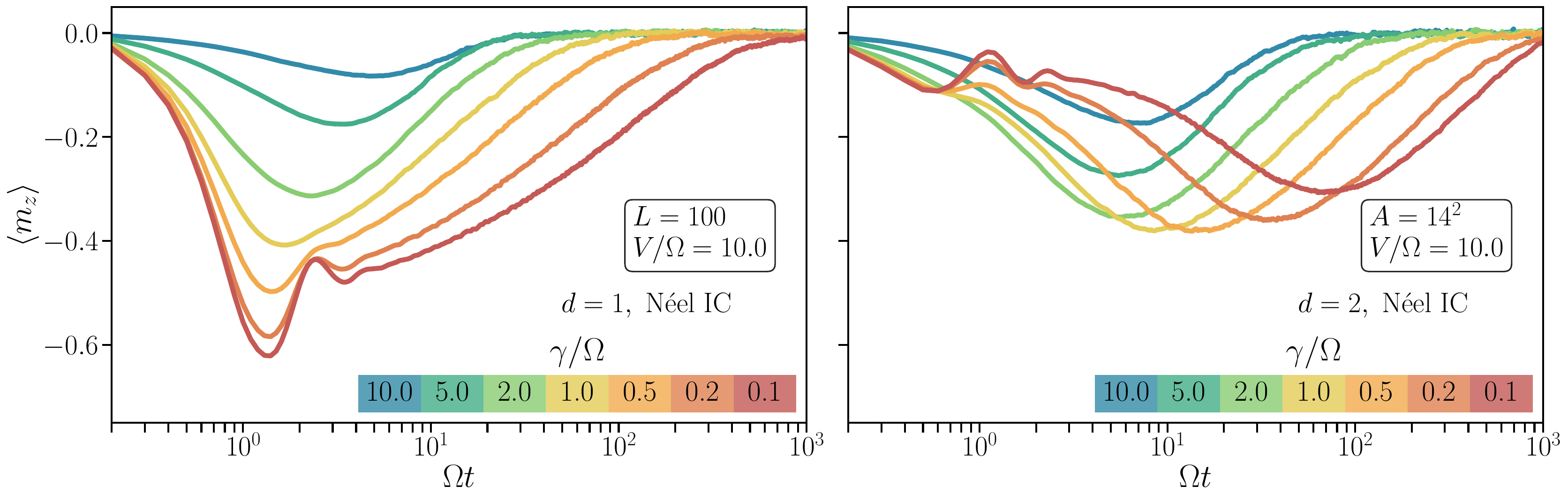}
    \caption{\textbf{Magnetisation relaxation at strong interaction and different dephasing strengths for the Néel initial state.}
    Dynamics for $V/\Omega=10.0$ in $d=1$ (left) and $d=2$ (right).
    The relaxation develops an increasingly deep intermediate-time minimum in $d=1$ as the dephasing is reduced, while in $d=2$ the minimum first becomes deeper, and then shallower again.
    For weak dissipation, pronounced coherent oscillations appear at early times. 
    Increasing dephasing suppresses these oscillations and reduces the overall deviation of the magnetisation from its initial (and stationary) value.}
    \label{fig:mag_rel_neel}
\end{figure*}

\begin{figure*} 
    \centering
    \includegraphics[width=0.99\linewidth]{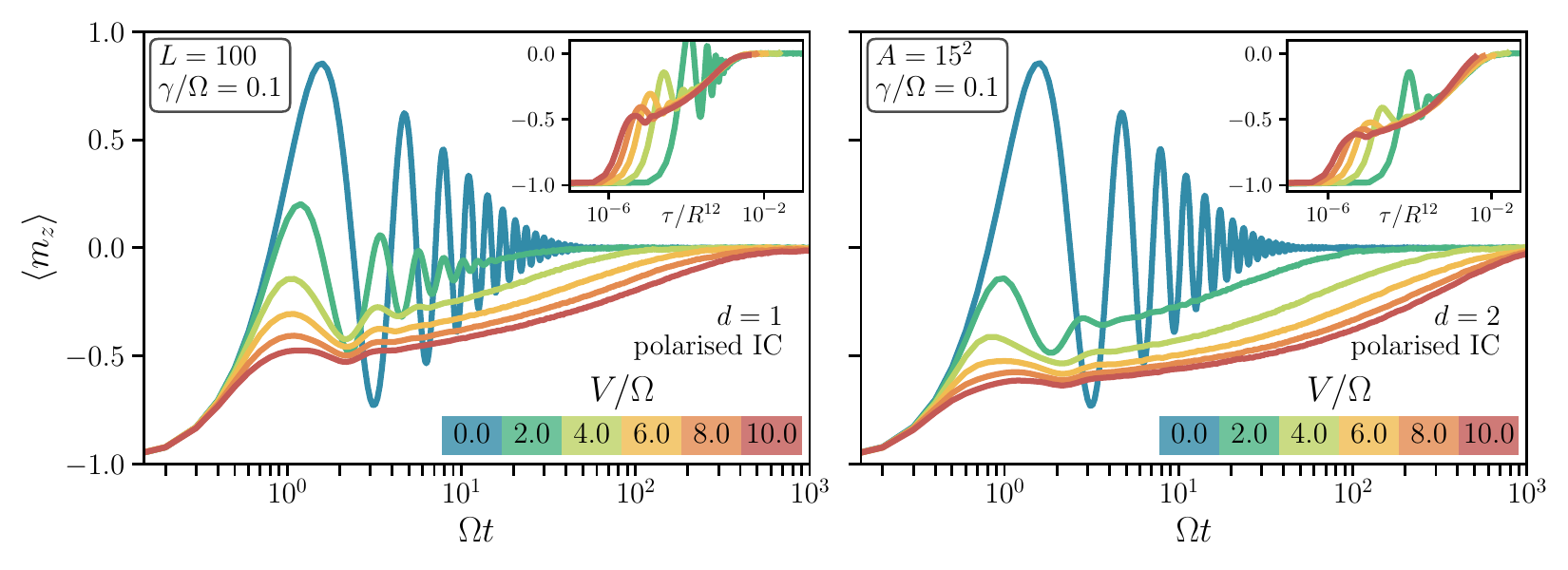}
    \caption{\textbf{Magnetisation relaxation at weak dephasing and different interactions for large systems}. Results are shown for $\gamma / \Omega=0.1$ in $d=1$ (left) and $d=2$ (right). For increasing $V / \Omega$, we can similarly observe signatures of an emerging plateau in both $d=1$ and $d=2$, as seen in Fig.~\ref{fig:mag_rel_taus} for decreasing dephasing at $V/\Omega=10.0$. In the insets, we rescale the relaxation curves by $R^{12}$. The collapse is very clear for $d=1$, while small deviations for $d=2$ at later times likely reflect the slight differences between TWA and the exact dynamics seen in Fig.~\ref{fig:bench1}(a). Note that the red curves in this plot and the red curves in Fig.~\ref{fig:mag_rel_taus}(a) represent the same data.}
    \label{fig:mag_rel2}
\end{figure*}

We consider different dephasing strengths $\gamma/\Omega$ in Fig.~\ref{fig:mag_rel_taus}(a) at fixed interaction $V/\Omega=10.0$ for $d=1$ (left) and $d=2$ (right) for the polarised initial state.
As seen on the left of Fig.~\ref{fig:mag_rel_taus}(a) for the $L=100$ chain, the relaxation approaches a plateau at $\langle m_z \rangle \simeq -0.48$, consistent with the hard-dimer constraint discussed in Sec.~\ref{sec:bench}.
We also observe the onset of an intermediate plateau in two dimensions (right hand side of Fig.~\ref{fig:mag_rel_taus}(a)), at magnetisation $\langle m_z\rangle \simeq -0.62$.
By analogy with the one-dimensional case, one might expect this corresponds to that of a hard-square lattice gas \cite{Ji_2011, Naik_2024, Ballar_Trigueros_2026}, the two-dimensional equivalent of the hard-dimer constraint, where the presence of an excitation suppresses the excitation of its four nearest neighbours. 
However, this would correspond to a magnetisation $\langle m_z \rangle \simeq -0.54$ \cite{Lesanovsky_2011}, which is substantially higher than the observed value.
This suggests that in two dimensions the dynamics in the coherent regime is subject to constraints extending beyond simple nearest-neighbour exclusion, indicating that coherent processes generate longer-ranged constraints that suppress spin flips over a larger surrounding region than in one dimension.

The relaxation curves exhibit distinct stages as the dynamics crosses over from coherent to dissipation-dominated behaviour. 
In this process, the magnetisation first approaches the plateau before slowly relaxing towards the stationary state.
To analyse these regimes more quantitatively, we extract two characteristic timescales from the dynamics, shown in Fig.~\ref{fig:mag_rel_taus}(b), (c).
The plateau relaxation time $\tau_{\text{plateau}}$, defined as the time to reach $\langle m_z(t) \rangle=-0.48$ in $d=1$ and $\langle m_z(t) \rangle=-0.62$ in $d=2$, scales as $\tau_{\text{plateau}}\sim \gamma$ in both dimensions.
The final relaxation time $\tau_{\text{final}}$, extracted by fitting $|m_z(t)| \sim e^{-t/\tau}$ in the regime $|m_z(t)| > 0.15$, obeys $\tau_{\text{final}} \sim 1/\gamma$ in both dimensions, indicating that the asymptotic relaxation is governed by the dissipative timescale. 
Consequently, decreasing $\gamma$ simultaneously accelerates the approach to the plateau via enhanced coherent dynamics, while slowing down the final relaxation towards the stationary state.
The $\tau_{\text{final}} \sim 1/\gamma$ scaling holds in the regime $\gamma \lesssim \Omega$, where the relaxation is rate-limited by dissipation. 
At the largest dephasing strengths $(\gamma/\Omega = 5,10)$, $\tau_{\text{final}}$ deviates from this behaviour and turns over, signalling the crossover to the strong-dephasing (quantum Zeno \cite{misra1977zeno, itano1990quantum}) regime: there, dephasing suppresses the coherent spin flips driving relaxation, and adiabatic elimination of the coherences gives an effective incoherent flip rate $\Gamma \sim \Omega^2/\gamma$ \cite{Lesanovsky_2013}, so that $\tau_{\text{final}} \sim \gamma/\Omega^2$. The two regimes cross near $\gamma \sim \Omega$, where the relaxation time is minimal.

We now turn to the investigation of the Néel initial condition in large systems, as shown in Fig.~\ref{fig:mag_rel_neel} for $L=100$ in $d=1$ (a) and $A=14^2$ in $d=2$ (b). 
In one dimension, the dynamics is qualitatively similar to that of the $L=10$ chain discussed in Sec.~\ref{sec:bench}.
In two dimensions, the time scales are more clearly separated.
Coherent oscillations dominate the early-time dynamics, while the minimum in the magnetisation develops only later.
The minimum reached in one dimension is noticeably lower than in two dimensions. 
The opposite trend is observed for the fully polarised initial state, where the emergent plateau occurs at a lower magnetisation in $d=2$. 
Starting from a lattice with no excitations, each excitation in two dimensions suppresses the creation of additional excitations on a larger number of neighbouring sites.
By contrast, starting from the Néel state, changing the excitation pattern leads to larger interaction shifts in two dimensions, which constrains the dynamics more strongly.

\begin{figure*}
    \centering
    \includegraphics[width=1.0\linewidth]{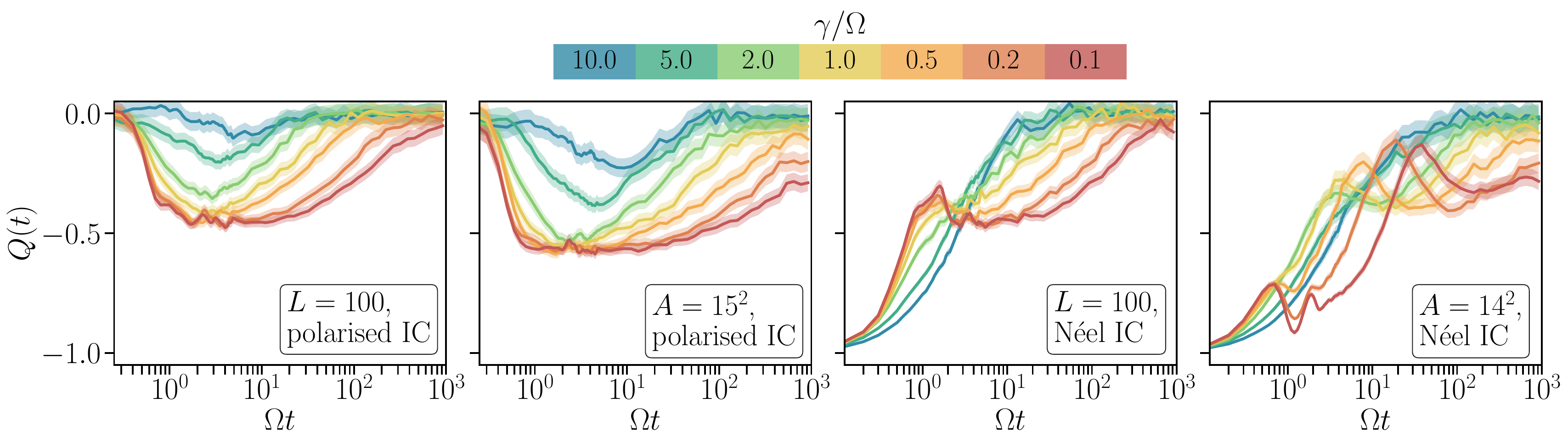}
    \caption{\textbf{Mandel $Q$-parameter for different dephasing strengths.}
    Time evolution of $Q(t)$ for the fully polarised initial state (left) and the Néel initial state (right) in $d=1,2$.
    Negative values of $Q(t)$ indicate suppressed magnetisation fluctuations compared to an uncorrelated lattice gas and therefore signal the presence of blockade-induced anticorrelations. 
    For the fully polarised initial state, decreasing the dephasing strength enhances the negative minimum of $Q(t)$ and prolongs the correlated regime.
    For the Néel initial state, $Q(t)$ starts close to its minimal value due to the initially ordered configuration and increases towards zero as this order is gradually lost during the relaxation. Error bars indicate the statistical uncertainty of $Q(t)$ estimated from the trajectory ensemble and correspond to 95$\%$ confidence interval; they are larger than for the mean magnetisation because $Q(t)$ is constructed from the magnetisation variance, a second moment.}
    \label{fig:c_spatial}
\end{figure*}

We now consider the effect of varying the interaction strength.
Fig.~\ref{fig:mag_rel2} shows the TWA relaxation curves for the polarised initial state at fixed dephasing $\gamma/\Omega = 0.1$, in $d=1$ (left) and $d=2$ (right) while varying the interaction strength.
Since kMC accurately describes the late-time relaxation even in the weakly dissipative regime, we characterise this slowdown by analysing the approach to the stationary state as a function of the effective interaction parameter $R$.
For van der Waals interactions ($\alpha = 6$), interaction-induced detuning shifts enter the classical rate equation \eqref{eq:rate-eq} through a squared term in the denominator of the rates \eqref{eq:rates}, leading to a suppression of spin flips that scales with the square of the interaction strength. 
In the strongly interacting regime $R\gg1$, the dominant contribution scales as $R^{2\alpha}$, which leads to an effective rate that scales as $\Gamma \sim 1/R^{12}$, implying a relaxation time $\propto R^{12}$.
Rescaling time by $R^{12}$ collapses the late-time dynamics onto a common curve, as shown in the insets of Fig.~\ref{fig:mag_rel2}.
This rescaling works similarly when considering the Néel initial state.

\subsubsection{Spatial correlations}

Spatial correlations provide additional insight into the emergence of collective behaviour during the dynamics.
To quantify these, we introduce the Mandel $Q$-parameter
\begin{equation}
Q(t) = \frac{\chi(t)}{\chi_{\mathrm{rnd}}(t)} - 1 ,
\label{eq:qval}
\end{equation}
where
\begin{equation}
\chi(t) = N\left(\langle m_z^2(t) \rangle - \langle m_z(t) \rangle^2\right)
\end{equation}
denotes the variance of the magnetisation.
As a reference value for the upcoming analysis, we consider the variance of a non-interacting lattice system with uncorrelated spins, given by
\begin{equation}
\chi_{\mathrm{rnd}}(t) = 1-\langle m_z(t)\rangle^2 .
\label{eq:xirand}
\end{equation}
The parameter $Q(t)$ therefore measures deviations from the behaviour of an uncorrelated random lattice gas: $Q(t)=0$ corresponds to uncorrelated spins, while $Q(t)\neq 0$ indicates the presence of correlations.
Fig.~\ref{fig:c_spatial} shows data for the $L=100$ spin chain and the $A=15^2 \,  (A=14^2)$ spin lattice at different  $\gamma/\Omega$ for the polarised and the  Néel initial states. 

For the fully polarised initial state in the strongly dissipative regime, $Q(t)$ remains close to zero for most of the evolution, with a small negative value developing at intermediate times.
In this limit, strong dephasing suppresses coherent processes and interactions do not generate significant correlations.
As the dephasing strength is reduced, $Q(t)$ develops a pronounced minimum with negative values during the relaxation process. 
This reflects the build-up of anticorrelations between excitations as the dynamics becomes dominated by blockade-induced kinetic constraints.
Correspondingly, the recovery of $Q(t)$ towards zero becomes progressively slower for weaker dephasing, reflecting the increasingly collective and constrained character of the relaxation. 
At late times, $Q(t)$ approaches zero as the system relaxes towards the stationary state.
This is consistent with Eqs.~\eqref{eq:qval}-\eqref{eq:xirand}, since the stationary state corresponds to an uncorrelated configuration of spins, where $Q(t)=0$.
Showing this more explicitly, we first take the magnetisation per site as $m_z=(1/N)\sum_i s_i^z$.
The local spin variable $s_i^z$ has the same mean value at every site in this case, $\langle s_i^z \rangle = \langle m_z \rangle$.
Since $s_i^z=\pm1$, we also have $(s_i^z)^2=1$.
The variance of the magnetisation then evaluates to 
$\langle m_z^2 \rangle - \langle m_z \rangle^2 = (1-\langle m_z \rangle^2)/N$.
This gives
$\chi(t) = N(\langle m_z^2 \rangle - \langle m_z \rangle^2) = 1 - \langle m_z \rangle^2 = \chi_{\mathrm{rnd}}(t)$.
The fluctuations $\chi(t)$ thus follow the random lattice gas result $\chi_{\mathrm{rnd}}(t)$, and therefore $Q = 0$. 
In the stationary state specifically, $\langle m_z \rangle = 0$ so that $\chi = 1$.

For the Néel initial state, $Q(t)$ shows a qualitatively different behaviour. 
Since excitations initially occupy only one sublattice, nearest-neighbour blockade constraints are already satisfied at $t=0$, leading to strongly suppressed magnetisation fluctuations. 
Accordingly, $Q(t)$ starts close to its minimal value and then increases towards zero as the initially ordered configuration is gradually lost during the relaxation.
The relaxation again depends on the dephasing strength. 
For strong dissipation, the increase of $Q(t)$ is smooth and monotonic, reflecting the incoherent decay of the initially ordered configuration. 
In the weakly dissipative regime, however, the relaxation proceeds through multiple stages.
The system first rapidly loses the initially ordered configuration. 
This is followed by a long-lived transient regime where $Q(t)$ remains close to $Q \simeq -0.5$, indicating that strong anticorrelations persist even after the initial Néel order has been lost. 
In this regime the evolution is constrained by the blockade-induced kinetic constraints, which restrict the allowed spin flips and thereby temporarily limit the accessible configurations.
At later times incoherent processes dominate and the system gradually approaches the uncorrelated stationary state, for which $Q=0$, consistent with the random lattice gas.
Overall, we also observe that weaker dissipation initially accelerates the approach to this intermediate flat region but slows the loss of correlations at later times, for both initial conditions.
This corroborates the observations in Fig.~\ref{fig:mag_rel_taus}(b), (c), where smaller values of $\gamma/\Omega$  lead to an earlier onset of the plateau in the magnetisation relaxation and a slower subsequent approach to the stationary state.
We analyse the behaviour of temporal correlation functions in App.~\ref{app:extradata}.

\section{Summary and outlook}
\label{sec:discussion}
In this work, we have investigated the relaxation dynamics of a dissipative Rydberg gas in one- and two-dimensional systems, focusing on the regime where coherent interactions and dissipation compete. 
To study this regime, we employed the TWA, which we first benchmarked against numerically exact simulations for small system sizes and then applied to large lattices beyond the reach of exact methods. 
Within this framework, we analysed the dynamics starting from two distinct initial states; a fully polarised state and a Néel state belonging to a quantum-scar manifold.

For the fully polarised initial state in one dimension, the dynamics evolves through an initial regime governed by resonant spin flips that increase the excitation density until the nearest-neighbour blockade restricts further spin flips, leading to the emergence of an intermediate plateau in the magnetisation. 
The late-time relaxation towards the stationary state is then governed by slow dissipative configuration changes, for both $d=1,2$.
Remarkably, a similar plateau is also observed in two dimensions in the weakly dissipative regime, which has not been found in the strongly dissipative limit.
This occurs at a magnetisation considerably lower than expected from a simple nearest-neighbour exclusion picture. 
This suggests the presence of effective constraints extending beyond nearest neighbours, whose role will be explored in more detail in future work.
Pronounced spatial anticorrelations, which decay to zero when the stationary state is reached further corroborate the presence of blockade-induced constraints. 
There are also characteristic time scales from the crossover between coherent and dissipative dynamics. 
Reducing the dissipation shortens the time required to reach the plateau in the coherent regime, while increasing the final relaxation times.

For the Néel initial state coherent oscillations dominate the early-time dynamics, however, by contrast the magnetisation develops a pronounced intermediate minimum rather than a plateau in the weakly dissipative regime.
During this regime strong anticorrelations persist even after the initial Néel order has been lost, reflecting the continued influence of blockade-induced constraints. 
At later times incoherent processes dominate and the system gradually approaches the uncorrelated stationary state.
The behaviour of the spatial correlations exhibits time scales similar to those found in the relaxation dynamics of the polarised initial state. 
In particular, a long-lived transient regime of anticorrelations emerges: for weaker dissipation this regime is reached more quickly, but the correlations also decay more slowly.

More broadly, our results demonstrate that the TWA offers a computationally efficient and scalable framework for probing the relaxation dynamics of open many-body systems in regimes that are currently inaccessible to exact methods, in particular large lattices and two spatial dimensions at weak to intermediate dissipation, where coherent driving and dissipation compete. 
The results are directly applicable to the large-scale Rydberg arrays now routinely realised in experiments, and provide a practical tool both for interpreting their nonequilibrium dynamics and for characterising the role of dissipation in Rydberg-based quantum simulation and computation, where decoherence and loss are intrinsic.

While TWA enables access to large system sizes and in higher dimensions, its limitations are apparent in regimes where kinetic constraints dominate.
At the same time, owing to the linear scaling of the TWA with system size, an extension of the present study to three-dimensional lattices is computationally feasible without much modification: only the lattice setup needs to be adapted. 
In three dimensions one would expect the blockade-induced constraints to be even more pronounced, as each excitation suppresses six nearest neighbours, although a quantitative benchmark of the method would be limited to small clusters accessible to exact methods.
In the one- and two-dimensional systems studied here, for the polarised initial state, only the onset of the intermediate magnetisation plateau associated with a nearest-neighbour excitation exclusion constraint is captured within TWA.
Extending the current approach, for instance via BBGKY hierarchies \cite{Pi_eiro_Orioli_2017} or cluster extensions of TWA \cite{Wurtz_2018} could allow for a more systematic incorporation of quantum correlations beyond the present treatment.
Tensor-network approaches could also provide an alternative route to explore this regime beyond phase-space methods.
This would be especially relevant for accurately describing the constraint-dominated regime in two dimensions, where the formation of the intermediate plateau could be studied at stronger interactions than in this work.

Lastly, we note that while the parameter regime explored here is, in principle, accessible in current Rydberg array experiments, the long-time dynamics considered in this work remains challenging to realise in practice.
Additionally, we have only considered dephasing as the source of dissipation. 
A finite Rydberg lifetime $\tau$ would introduce an additional decay channel at rate $1/\tau$, setting an observation window for this dynamics. 
The constraint-dominated features reported here are most pronounced around $\Omega t \sim 10^1$ and thus remain accessible provided $\Omega\tau \gtrsim 10^1$, whereas the full relaxation to the stationary state at $\Omega t \sim 10^3$ is more demanding. At longer times, decay competes with excitations, and creates a correlated steady state, as shown in Ref.~\cite{Lesanovsky_2013} in the strongly dissipative limit. 
The observation of the slow, constraint-dominated relaxation requires coherence and stability over extended times, whereas dissipation in typical setups leads to heating and loss processes that can obscure these effects.
Achieving such regimes would therefore require dissipative quantum platforms that remain stable at long times. 
Promising candidates in this direction include recently proposed Rydberg-ion systems, where engineered dissipation and longer coherence times may enable access to this dynamics \cite{martins2026quantumsimulationrydbergions}.
\section*{Code and Data availability}
The numerical data supporting the findings of this work is available at \cite{zenodo_data}, and the simulation code used to generate the data can be found at \cite{zenodo_software}.
\begin{acknowledgments}
We thank M. Cech for fruitful discussions, comments on the manuscript and for helpful guidance with quantum jump Monte Carlo simulations.
This work is supported by the ERC grant OPEN-2QS (Grant No. 101164443).
We acknowledge funding from the Deutsche Forschungsgemeinschaft (DFG, German Research Foundation) through the Research Unit FOR 5413/1, Grant No. 465199066 and through the Research Unit FOR 5522/1, Grant No. 499180199. We also acknowledge
funding through JST-DFG 2024: Japanese-German
Joint Call for Proposals on “Quantum Technologies”
(Japan-JST-DFG-ASPIRE 2024) under DFG Grant No.
55456179.
The authors acknowledge support by the state of Baden-Württemberg through bwHPC and the DFG through grant no INST 40/575-1 FUGG (JUSTUS 2 cluster).
ChatGPT (OpenAI, GPT-3) was used for minor improvements to figure presentation and language editing.
\end{acknowledgments}
\appendix 
\section{Truncated Wigner approach to dissipative spins}
\label{app:TWA}
In this appendix, we summarise the necessary steps for the derivation of the TWA equations used in the main text.
First, we briefly outline the general Lindblad-to-TWA procedure and then apply it specifically to the spin model considered in this work.
The presentation follows from the Keldysh path-integral formulation of Lindblad dynamics and its semiclassical truncation, as detailed in Ref.~\cite{Hosseinabadi:2025xbk, Sieberer_2016}.

\subsection{Keldysh path-integral formulation}

The Keldysh path integral is defined on a closed time contour $\mathcal{C}$ running from $t_i$ to $t_f$ and back to $t_i$ \cite{keldysh2024diagram, schwinger1961brownian}. 
In the absence of dissipation, for the fields $\psi^\pm$ on the forward and backward branches, the generating functional reads 
\begin{equation}
    Z = \int \mathcal D[\psi^{\pm}]\, P_0[\psi_0^\pm]\, e^{iS_{\text{Keldysh}}[\psi^{\pm}]},
\end{equation}
where $\mathcal{D}[\psi^{\pm}] \equiv \mathcal{D}[\psi^+,\psi^-]$ is the functional integration measure over the forward and backward fields.
The Keldysh action is given by
\begin{equation}
S_{\text{Keldysh}} = \int_{t_i}^{t_f}dt \left[\mathcal{L}(\psi^+)  - \mathcal{L}(\psi^-)\right],
\label{eq:coh-keldysh}
\end{equation}
where $\mathcal{L}(\psi^{\pm})$ denotes the Lagrangian density of the system.
Expectation values are obtained from
\begin{equation}
\langle O(t) \rangle = \int \mathcal{D}[\psi^{\pm}]\, P_0[\psi_0^{\pm}]\, O(t)\, e^{i S_{\text{Keldysh}}},
\end{equation}
where $P_0[\psi_0^{\pm}]\equiv P_0[\psi_0^+,\psi_0^-]$ encodes the phase-space representation of the initial density matrix.
For an open system described by a Lindblad master equation \eqref{eq:Lindblad}, the action receives an additional dissipative contribution,
\begin{equation}
S = S_{\text{Keldysh}} + S_{\mathrm{diss}},
\end{equation}
with \cite{Sieberer_2016}
\begin{equation}
S_{\mathrm{diss}}=- i \sum_i \frac{\gamma_i}{2} \int dt \left(2 L_i^+ \bar L_i^- - L_i^+ \bar L_i^+ - L_i^- \bar L_i^-\right),
\label{eq:sdiss}
\end{equation}
where $L_i^\pm$ denote the jump operators evaluated on the forward and backward branches of the contour. 
Here and throughout, an overbar on a single field or operator, such as $\bar{L}_i$ denotes complex conjugation. 
We use this instead of $L^*$ to avoid the cluttering of symbols with the Keldysh superscripts $\pm$. 
More generally, Eq.~\eqref{eq:sdiss} follows from a bosonic coherent-state path integral. 
For an $\mathrm{SU}(2)$ phase space, however, an additional subtlety arises: the jump operators must be combined using the star product rather than ordinary multiplication. 
In the case of pure dephasing, which is what we consider in this case, this distinction is inconsequential, whereas for decay it leads to corrections. 
We will discuss this point in detail in future work \cite{noel2026su2keldysh}. 
\subsection{Keldysh rotation}

Introducing the Keldysh rotation, the fields can be written in terms of ``classical'' and ``quantum'' components, given by
\begin{equation}
\psi^c = \tfrac12(\psi^+ + \psi^-), \qquad \psi^q = \psi^+ - \psi^-.
\end{equation}
In this basis, the action admits an expansion in powers of $\psi^q$, where the linear term follows by expanding the coherent contribution \eqref{eq:coh-keldysh} in $\psi^\pm=\psi^c\pm\tfrac12\psi^q$,
\begin{equation}
\mathcal{L}(\psi^c+\tfrac12\psi^q)-\mathcal{L}(\psi^c-\tfrac12\psi^q)=\psi^q_\alpha\,\frac{\delta \mathcal{L}}{\delta \psi_\alpha}\Big|_{\psi^c}+\mathcal{O}\!\big((\psi^q)^3\big),
\end{equation}
so that
\begin{equation}
S_{\text{Keldysh}} =\int dt\, \psi^q_\alpha\,\mathcal{E}_\alpha[\psi^c] +\mathcal{O}\!\big((\psi^q)^3\big), 
\end{equation}
with 
\begin{equation}
    \mathcal{E}_\alpha[\psi^c]\equiv \frac{\delta \mathcal{L}}{\delta \psi_\alpha}\Big|_{\psi^c}.
\end{equation}
Here, the index $\alpha$ labels internal and site degrees of freedom, for instance, for spins, one has $\alpha = (i,\mu)$ with $\mu = x,y,z$, where $i$ is the site index, and $\mu$ are for the three spin components
The quadratic term originates from the dissipative contribution \eqref{eq:sdiss}. 
Writing
\begin{equation}
L_i^\pm = L_i^c \pm \tfrac12 L_i^q,\quad\bar L_i^\pm = \bar L_i^c \pm \tfrac12 \bar L_i^q,
\end{equation}
one finds
\begin{equation}
2 L_i^+ \bar L_i^- - L_i^+ \bar L_i^+ - L_i^- \bar L_i^-=- \tfrac12 L_i^q \bar L_i^q .
\end{equation}
Substituting into \eqref{eq:sdiss} gives
\begin{equation}
S_{\mathrm{diss}}=\frac{i}{4}\sum_i \gamma_i\int dt\, L_i^q \bar L_i^q.
\end{equation}
Altogether, we have
\begin{equation}
S[\psi^c,\psi^q]=\int dt\,\Big[\psi^q_\alpha\,\mathcal{E}_\alpha[\psi^c]+\frac{i}{2}\,\psi^q_\alpha\, D_{\alpha\beta}[\psi^c]\,\psi^q_\beta+\mathcal{O}\!\big((\psi^q)^3\big)\Big].
\label{eq:generic_keldysh_structure}
\end{equation} 
The kernel $D_{\alpha\beta}[\psi^c]$ is determined by the dissipative term \eqref{eq:sdiss} through the quadratic form $L_i^q \bar L_i^q$.
Higher-order terms in $\psi^q$ represent further quantum corrections.

\subsection{Semiclassical approximation and emergence of the stochastic equations}

The truncated Wigner approximation consists of truncating the action \eqref{eq:generic_keldysh_structure} at quadratic order in the quantum fields $\psi^q$.
The weight $e^{iS}$ in the path integral now reads 
\begin{equation}
e^{iS}=\exp\!\left(
i\int dt\,\psi^q_\alpha \mathcal E_\alpha[\psi^c]\right)\,\exp\!\left(i\frac{i}{2}\int dt\,\psi^q_\alpha D_{\alpha\beta}\psi^q_\beta\right).
\end{equation}
The quadratic term can be represented by a Gaussian identity.
Introducing auxiliary complex Gaussian fields $\xi_\alpha(t)$,
\begin{align}
&\exp\!\left[-\frac{1}{2}\int dt\,\psi^q_\alpha D_{\alpha\beta}\psi^q_\beta\right]\propto\nonumber\\
&\int \mathcal D[\xi]\,\exp\!\left[-\frac{1}{2}\int dt\,\xi_\alpha (D^{-1})_{\alpha\beta}\xi_\beta\right.\left.+\, i\int dt\,\xi_\alpha \psi^q_\alpha\right],
\label{eq:hub-strat}
\end{align}
the action becomes linear in $\psi^q$,
\begin{equation}
S=\int dt\,\psi^q_\alpha\big(\mathcal{E}_\alpha[\psi^c]-\xi_\alpha\big)+S_\xi,
\end{equation}
with 
\begin{equation}
S_\xi=-\frac{1}{2}\int dt\, \xi_\alpha (D^{-1})_{\alpha\beta}\xi_\beta,
\end{equation}
which implies
\begin{equation}
\overline{\xi_\alpha(t)\bar\xi_\beta(t')}=D_{\alpha\beta}\,\delta(t-t').
\end{equation}
The functional integral over $\psi^q$ can thus be performed explicitly,
\begin{align}
\int \mathcal D[\psi^q]\,
\exp\!\left[
i\int dt\, \psi^q_\alpha(t)\,
\big(\mathcal{E}_\alpha[\psi^c](t)-\xi_\alpha(t)\big)
\right]
\nonumber\\
\propto\;
\delta\!\Big[\mathcal{E}_\alpha[\psi^c]-\xi_\alpha\Big].
\end{align}
where $\delta\!\Big[\mathcal{E}_\alpha[\psi^c]-\xi_\alpha\Big]$ denotes a functional delta, i.e.\ $\delta[F]\equiv\prod_{t,\alpha}\delta(F_\alpha(t))$.
The generating functional therefore reduces to
\begin{equation}
Z=\int \mathcal D[\psi^c] \mathcal D[\xi]\,P_0[\psi_0^c]\,\delta\!\big(\mathcal{E}_\alpha[\psi^c]-\xi_\alpha\big)\,e^{i S_\xi},
\label{eq:delta_functional}
\end{equation}
which is the functional representation of a stochastic process.
The delta functional enforces the stochastic equations
\begin{equation}
\mathcal{E}_\alpha[\psi^c]=\xi_\alpha(t).
\label{eq:constraint}
\end{equation}
In the absence of dissipation ($D_{\alpha\beta}=0$),
Eq.~\eqref{eq:constraint} reduces to the classical equations
\begin{equation}
\mathcal{E}_\alpha[\psi^c]=0.
\end{equation}
These equations can be written in Hamiltonian form as
\begin{equation}
\dot\psi_\alpha = \{\psi_\alpha,H\}_p.
\end{equation}
This Poisson structure corresponds to the classical limit of the quantum commutator $\{A,B\}_p \longleftrightarrow -i[\hat A,\hat B]$ (Dirac correspondence).

For an open system, the kernel $D_{\alpha\beta}$ originates from the quadratic form $L_i^q \bar L_i^q$ in the dissipative action \eqref{eq:sdiss}. 
In a phase-space formulation, the quantum fields $\psi^q$ play the role of response fields and generate infinitesimal canonical transformations of the classical variables.
Consequently, to leading order in $\psi^q$, the difference of any functional $F[\psi]$ between the two contour branches is generated by the structure
\begin{equation}
F^q \equiv F[\psi^+]-F[\psi^-]=\psi^q_\alpha\,\{\psi_\alpha,F\}_p+\mathcal O\!\big((\psi^q)^2\big),
\label{eq:Fq_poisson}
\end{equation}
so that $L_i^q=\psi^q_\alpha\{\psi_\alpha,L_i\}_p+\cdots$ and similarly for $\bar L_i^q$.
Substituting this into $S_{\mathrm{diss}}\propto \int dt\,L_i^q\bar L_i^q$ yields a quadratic form
in $\psi^q$ with kernel
\begin{equation}
D_{\alpha\beta}[\psi^c]
\propto
\sum_i \gamma_i\,
\{\psi_\alpha,L_i\}_p\,
\{\psi_\beta,\bar L_i\}_p,
\label{eq:D_from_L}
\end{equation}
which is positive semidefinite by construction.
After the Hubbard--Stratonovich decoupling \eqref{eq:hub-strat}, the stochastic equations $\mathcal{E}_\alpha[\psi^c]=\xi_\alpha(t)$ involve a Gaussian noise $\xi_\alpha$ with covariance $D_{\alpha\beta}$. 
Since $D_{\alpha\beta}$ factorises in terms of jump-operator brackets as in \eqref{eq:D_from_L}, it is convenient to parameterise $\xi_\alpha$ in terms of independent Gaussian noises $\xi_i(t)$ associated with each jump operator $
\xi_\alpha(t) = \sum_i \sqrt{\gamma_i}\, \{\psi_\alpha,L_i\}_p\, \xi_i(t).$
As the Lindblad dissipator generates both deterministic dissipative drift terms and stochastic fluctuations, it is convenient to combine these contributions by introducing auxiliary fields
\begin{equation}
\Phi_i=\gamma_i L_i+\xi_i(t),
\end{equation}
where $\gamma_i L_i$ produces the deterministic dissipative contribution and $\xi_i(t)$ represents the Gaussian noise associated with the jump process.
Thus, the stochastic contribution can be expressed as
\begin{equation}
\xi_\alpha(t)=-\frac{i}{2}\sum_i\left(\{\psi_\alpha,\bar L_i\}_p\,\Phi_i-\bar\Phi_i\,\{L_i,\psi_\alpha\}_p\right),
\label{eq:Xi_def}
\end{equation}
in the more general equations of motion
\begin{equation}
\dot\psi_\alpha = \{\psi_\alpha,H\}_p + \xi_{\alpha}(t),
\end{equation}
which reproduces the stochastic equations \eqref{eq:twa-stoch} in the main text.

We note that the resulting dynamics can equivalently be interpreted as the semiclassical limit of a quantum Langevin equation \cite{gardiner2004handbook}.

\subsection{Application to the long-range interacting, dissipative Rydberg gas}

For spin-$1/2$ degrees of freedom, operators $\hat\sigma_k^\alpha$ are replaced by classical vectors
\begin{equation}
\mathbf{s}_i = (s_i^x, s_i^y, s_i^z),
\end{equation}
with Poisson algebra
\begin{equation}
\{s_i^\alpha, O\}_p = 2 \sum_{\beta\gamma} \epsilon_{\alpha\beta\gamma} \frac{\partial O}{\partial s_i^\beta} s_i^\gamma,
\end{equation}
where $\epsilon_{\alpha\beta\gamma}$ is the totally antisymmetric Levi--Civita tensor, with $\epsilon_{xyz}=+1$.
In particular, the fundamental Poisson brackets reproduce the classical $\mathrm{SU}(2)$ algebra,
\begin{equation}
\{s_i^\alpha, s_j^\beta\}_p=2\,\delta_{ij}\,\epsilon_{\alpha\beta\gamma}\,s_i^\gamma.
\end{equation}
The classical Hamiltonian is
\begin{equation}
H=\Omega \sum_i s_i^x+\sum_{i<j}V_{ij}\left(\frac{1+s_i^z}{2}\right)\left(\frac{1+s_j^z}{2}\right),
\end{equation}
with long-range interaction \eqref{eq:longrange}, and the classical dephasing jump operator is
\begin{equation}
L_{i,\gamma}=\sqrt{\gamma}\,\frac{1+s_i^z}{2}.
\end{equation}

The coherent equations of motion follow from $\dot s_i^\alpha = \{s_i^\alpha, H\}_p$. 
For the driving term, one obtains
\begin{equation}
\left.\dot s_i^x\right|_{\Omega} = 0, \quad
\left.\dot s_i^y\right|_{\Omega} = -2\Omega s_i^z, \quad
\left.\dot s_i^z\right|_{\Omega} = +2\Omega s_i^y,
\end{equation}
while for the interaction term, the contributions are
\begin{equation}
\left.\dot s_i^x\right|_{V} = - V_i^{\mathrm{eff}} s_i^y, \quad
\left.\dot s_i^y\right|_{V} = + V_i^{\mathrm{eff}} s_i^x, \quad
\left.\dot s_i^z\right|_{V} = 0,
\end{equation}
with $V_i^{\mathrm{eff}}$ given by Eq.~\eqref{eq:veff}.
For the local dephasing, the stochastic contributions are
\begin{equation}
\left.\dot s_i^x\right|_{\gamma} = +2 \eta_i s_i^y, \quad
\left.\dot s_i^y\right|_{\gamma} = -2 \eta_i s_i^x, \quad
\left.\dot s_i^z\right|_{\gamma} = 0,
\end{equation}
where the real Gaussian noise satisfies Eq.~\eqref{eq:realnoise}.

\section{Numerical implementation}
\label{app:numerics}
\subsection{Stochastic trajectories of spins}
In this appendix, we summarise the numerical implementation used to simulate the dissipative spin dynamics discussed in the main text.
The dynamics of each site $i$ in our lattice is represented by a classical spin vector 
$\mathbf{s}_i(t) = (s^x_{i}, s^y_{i}, s^z_{i}) \in \mathbb{R}^3$ of fixed length $|\mathbf{s}_i| = r_0$, whose orientation evolves due to coherent precession and stochastic noise.
The time evolution is governed by a Stratonovich stochastic differential equation of the form
\begin{equation}
d\mathbf{s}_i 
  = \mathbf{f}_i(\{\mathbf{s}_j\})\,dt
  + \mathbf{g}_{i}(\mathbf{s}_i)\circ dW_{i}(t),
\label{eq:sde_stratonovich}
\end{equation}
where the symbol ``$\circ$'' denotes the Stratonovich interpretation. 
The deterministic drift is given by
\begin{equation}
\mathbf{f}_i(\{\mathbf{s}_j\}) =
\begin{pmatrix}
- V_i^{\mathrm{eff}} s_i^y \\
V_i^{\mathrm{eff}} s_i^x - 2\Omega s_i^z \\
2\Omega s_i^y
\end{pmatrix},
\end{equation}
which describes coherent precession in the effective field
$(2\Omega,0,V_i^{\mathrm{eff}})$, where $V_i^{\mathrm{eff}} $ is the interaction field generated by the surrounding spins, while the noise vector is
\begin{equation}
\mathbf{g}_i(\mathbf{s}_i) =
\begin{pmatrix}
2 \sqrt{\gamma} s_i^y \\
-2 \sqrt{\gamma} s_i^x \\
0
\end{pmatrix},
\end{equation}
generating stochastic rotations about the $z$-axis.
The increments $dW_i(t)$ are independent Wiener processes satisfying
\begin{equation}
\langle dW_i(t)\, dW_j(t) \rangle = \delta_{ij}\, dt.
\end{equation}
Both the deterministic and stochastic contributions are constructed such that they are tangent to the Bloch sphere, i.e.
\begin{equation}
\mathbf{s}_i\!\cdot\!\mathbf{f}_i = 0, \qquad \mathbf{s}_i\!\cdot\!\mathbf{g}_{i} = 0,
\label{eq:tangent_condition}
\end{equation}
for all $i$.
This property follows from the fact that both the coherent mean-field evolution and the Lindblad jump operators generate infinitesimal $\mathrm{SU}(2)$ rotations of the local spin vector.
As a consequence, the dynamics can be written entirely in terms of an effective angular velocity vector $\boldsymbol{\omega}_i(\mathbf{s})$ via
\begin{equation}
d \mathbf{s}_i = (\boldsymbol{\omega}_i(\mathbf{s}) \times \mathbf{s}_i) \, dt,
\label{eq:so3_flow}
\end{equation}
where $\boldsymbol{\omega}_i$ is the sum of a deterministic contribution $\boldsymbol{\omega}_i^{(\mathrm{det})} = (2\Omega,\,0,\,V_i^{\mathrm{eff}})$ and a stochastic contribution $\boldsymbol{\omega}_i^{(\mathrm{noise})}=-2\eta_i(t)\hat{z}$, where $\eta_i(t)$ denotes Gaussian white noise in Eq.~\ref{eq:realnoise}.
Eq.~\eqref{eq:so3_flow} implies that each realisation of the dynamics is a trajectory on the $S^2$ sphere of radius $r_0$:
\begin{equation}
d|\mathbf{s}_i|^2 
  = 2\,\mathbf{s}_i\!\cdot\!d\mathbf{s}_i
  = 2\,\mathbf{s}_i\!\cdot\!(\boldsymbol{\omega}_i\times\mathbf{s}_i) \, dt = 0.
\end{equation}
Thus, in exact arithmetic, every trajectory preserves the spin length $|\mathbf{s}_i|$.

\subsection{Rotation-based Stratonovich integrator}
In numerical implementations, conventional SDE solvers (such as Euler–Maruyama or Heun-type schemes) perform linear additive updates,
\begin{equation}
\mathbf{s}_i^{n+1}
  \approx \mathbf{s}_i^{n}
  + \mathbf{f}_i^{n}\,\Delta t
  + \mathbf{g}_i^{n}\,\Delta W_{i}
\end{equation}
rather than finite rotations on the sphere. 
As a consequence, the discrete update is not an orthogonal transformation and does not strictly preserve the constraint $|\mathbf{s}_i| = r_0$. 
Although the per-step deviation is small, it accumulates over the long simulation times considered in the main text and leads to a gradual drift of the spin norm.
A simple remedy is to project the updated spin back onto the sphere after each step, so that $\mathbf{s}_i^{n+1} \;\rightarrow\; r_0\,\mathbf{s}_i^{n+1}/|\mathbf{s}_i^{n+1}|,$
which restores the constraint at the discrete level. However, this projection slightly modifies the stochastic flow.
This motivates the use of structure-preserving integrators that evolve each timestep as a finite rotation $R_i(\Delta t)\in \mathrm{SO}(3)$ on the Bloch sphere, thereby maintaining the constraint intrinsically.

\begin{figure*}[htb!]
    \centering
     \includegraphics[width=1.0\linewidth]{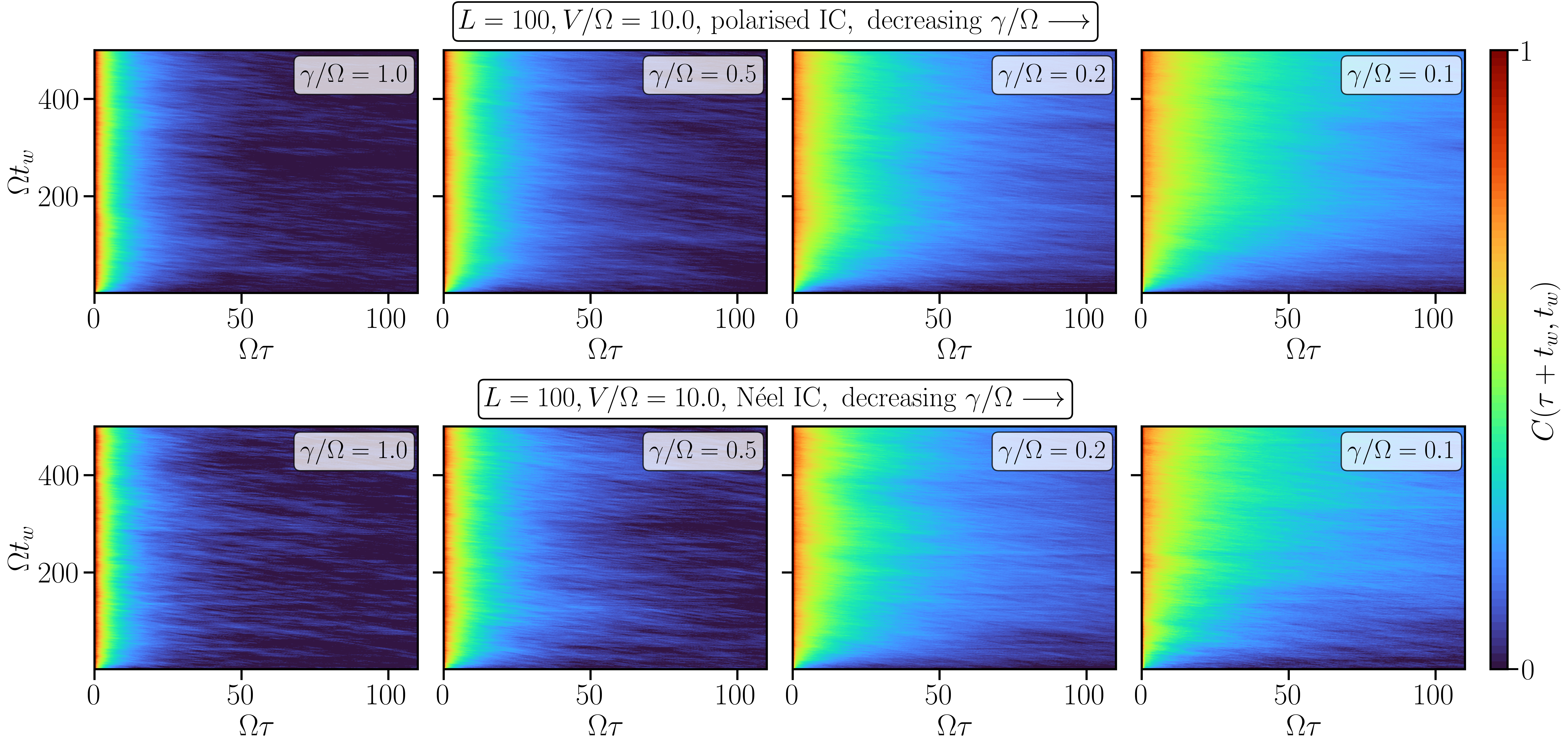}
    \caption{\textbf{Connected two-time autocorrelation for a chain of $L=100$ spins with interaction strength $V/\Omega=10$.}
The top row shows results for the fully polarised initial state, while the bottom row corresponds to the Néel initial state. 
The horizontal axis shows the waiting time $t_w$ and the vertical axis the time difference $\tau = t - t_w$. 
The colour scale indicates the value of the connected autocorrelation. As the dissipation strength decreases, correlations persist for longer times and the decay becomes progressively slower, reflecting the increasing influence of coherent dynamics.}
    \label{fig:c_two_time_1D}
\end{figure*}
\begin{figure*}[htb!]
    \centering
    \includegraphics[width=1.0\linewidth]{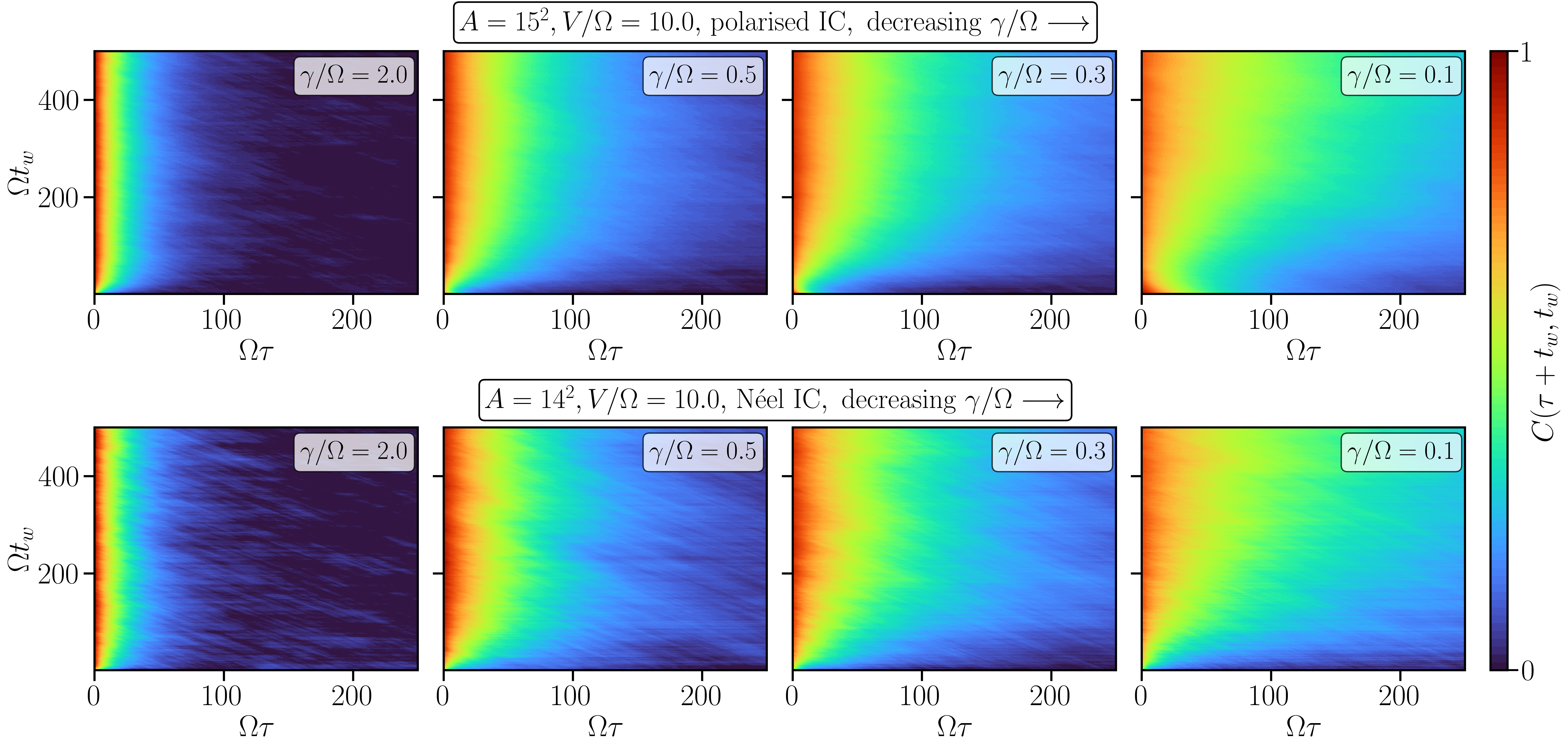}
    \caption{\textbf{Connected two-time autocorrelation for a lattice of $L=15^2 (14^2)$ spins with interaction strength $V/\Omega=10$.}
The top row shows results for the fully polarised initial state, while the bottom row corresponds to the Néel initial state. As the dissipation strength decreases, correlations persist for longer times and the decay becomes progressively slower, reflecting the increasing influence of coherent dynamics.}
    \label{fig:c_two_time}
\end{figure*}

Equation~\eqref{eq:so3_flow} generates a flow on the Lie group $\mathrm{SO}(3)$ and therefore preserves $|\mathbf{s}_i|$ exactly \cite{malham2007stochasticliegroupintegrators}.
To maintain this structure at the discrete level, we evolve each timestep as a composition of finite rotations.
We employ a symmetric Strang-type splitting scheme for the Stratonovich SDE in the spirit of \cite{ninomiya2008weak}, consisting of: (i) a deterministic half-step rotation, (ii) stochastic rotations generated by the Wiener increments, and (iii) a second deterministic half-step rotation evaluated at the updated configuration.
Here, the drift in the second half-step is recomputed from the intermediate configuration obtained after the stochastic rotations.
Letting $\mathbf{s}_i^{n}$ denote the spin at time $t_n$, the full timestep therefore corresponds to the composition
\begin{equation}
\mathbf{s}_i^{n+1} = R_{\mathrm{det}}\!\left(\tfrac{\Delta t}{2}\right)
\,R_{\mathrm{noise}}(\Delta W_i)\,
R_{\mathrm{det}}\!\left(\tfrac{\Delta t}{2}\right)
\,\mathbf{s}_i^{n},
\end{equation}
which preserves $|\mathbf{s}_i|$ up to machine precision.
Since the stochastic rotations are applied symmetrically between two deterministic half-steps, the resulting composition constitutes a symmetric Strang splitting of the Stratonovich SDE on $\mathrm{SO}(3)$. 
As rotations on $\mathrm{SO}(3)$ do not commute in general, the symmetric placement of the stochastic rotation between the two deterministic half-steps is important for maintaining consistency with Stratonovich calculus.

For the deterministic part, the drift $\mathbf{f}_i$ (using $\mathbf{f}_i=\boldsymbol{\omega}_i\times\mathbf{s}_i$) determines an instantaneous angular velocity 
\begin{equation}
\boldsymbol{\omega}_i = \frac{\mathbf{s}_i \times \mathbf{f}_i}{|\mathbf{s}_i|^2},
\end{equation}
and the spin is rotated by an angle $\tfrac{1}{2}|\boldsymbol{\omega}_i|\,\Delta t$ about the axis $\boldsymbol{\omega}_i/|\boldsymbol{\omega}_i|$.
The stochastic part is implemented as exact finite rotations about the corresponding axes. 
For example, dephasing generates rotations about $\hat{z}$ with angle
\begin{equation}
\theta_i^{(z)} = 2\sqrt{\gamma}\,\Delta W_i
\end{equation}
where $\Delta W_i$ are Wiener increments.

\section{Temporal correlations}
\label{app:extradata}

To probe the temporal correlations underlying the relaxation dynamics, we analyse the connected two-time autocorrelation function, 
\begin{equation}
C(t_w+\tau,t_w) = \frac{1}{N}\sum_i \langle s_i^z(t_w+\tau)s_i^z(t_w) \rangle_c,
\end{equation}
where $\tau \ge 0$ denotes the time difference after the waiting time $t_w$, i.e., $\tau = t - t_w$.
This is shown for decreasing dephasing strength for the polarised initial state (top row) as well as the Néel initial state (bottom row) for the $d=1$ spin chain in Fig.~\ref{fig:c_two_time_1D} and for the $d=2$ lattice in Fig.~\ref{fig:c_two_time}.

The contour plots reveal a similar behaviour of the relaxation on the waiting time $t_w$ in both $d=1$ and $d=2$, and for both initial states.
For strong dephasing the autocorrelation decays rapidly in the time difference $\tau$, and the contour lines are nearly vertical, indicating that the dynamics quickly becomes stationary and loses memory of the initial configuration. 
As the dephasing strength is reduced, correlations persist over increasingly longer time differences and a pronounced dependence on the waiting time $t_w$ emerges. 
This appears as a bending of the contour lines at intermediate $t_w$, reflecting that the effective relaxation rate evolves during the approach to the stationary state.
However, for a fixed waiting time $t_w$, the decay in $\tau$ remains exponential throughout the evolution. 
The nontrivial connected two-time correlations arise because the interacting dynamics retains memory of earlier spin configurations during relaxation. 
Since the spin-flip rates depend on the local excitation pattern, the configuration at time $t_w$ influences the subsequent evolution, leading to a waiting-time dependent autocorrelation until the stationary state is reached.
While the same qualitative behaviour is observed in one and two dimensions, the regime of pronounced waiting-time dependence extends over longer times in $d=2$, indicating slower relaxation and more persistent temporal correlations.

\bibliography{main_text/main}
\end{document}